\documentclass[twocolumn, journal]{IEEEtran}
\usepackage{color}
\usepackage{graphicx}
\usepackage{epstopdf}
\usepackage{amsmath}
\usepackage{amssymb}
\usepackage{algorithm}
\usepackage{algorithmic}

\usepackage{enumerate}
\usepackage{amsmath}
\usepackage{multirow}
\usepackage{booktabs}
\usepackage{array}
\usepackage{amsthm}
\usepackage{stfloats}
\usepackage{caption}
\usepackage{subfigure}
\usepackage{bm}
\usepackage{booktabs}
\usepackage{setspace}
\usepackage{gensymb}
\usepackage{url}

\allowdisplaybreaks[4]
\title{Optimization for Reflection and Transmission Dual-Functional Active RIS-Assisted Systems}

\author{Yanan Ma, Ming Li,~\IEEEmembership{Senior Member,~IEEE}, Yang Liu,~\IEEEmembership{Member,~IEEE},\\
 Qingqing Wu,~\IEEEmembership{Member,~IEEE}, and Qian Liu,~\IEEEmembership{Member,~IEEE}
\thanks{Yanan Ma, Ming Li and Yang Liu, are with the School of Information and Communication Engineering, Dalian University of Technology, Dalian, Liaoning 116024, China, (e-mail: mayanan@mail.dlut.edu.cn, mli@dlut.edu.cn, yangliu\_613@dlut.edu.cn).}
\thanks{Qingqing Wu is with the State Key Laboratory of Internet of Things for
Smart City, University of Macau, Macau 999078, China (e-mail: qingqingwu@um.edu.mo).}
\thanks{Qian Liu is with the School of Computer Science and Technology, Dalian University of Technology, Dalian 116024, China (e-mail: qianliu@dlut.edu.cn).}}
\begin{document}

\maketitle

\pagestyle{empty}  % no page number for the second and the later pages
\thispagestyle{empty} % no page number for the first page

\begin{abstract}
Reconfigurable intelligent surface (RIS) has been deemed as one of potential components of future wireless communication systems because it can adaptively manipulate the wireless propagation environment with low-cost passive devices. However, due to the severe double path loss, the traditional passive RIS can provide sufficient gain only when receivers are very close to the RIS. Moreover, RIS cannot provide signal coverage for the receivers at the back side of it. To address these drawbacks in practical implementation, we introduce a novel reflection and transmission dual-functional active RIS architecture in this paper, which can simultaneously realize reflection and transmission functionalities with active signal amplification to significantly extend signal coverage and enhance the quality-of-service (QoS) of all users. The problem of joint transmit beamforming and dual-functional active RIS design is investigated in RIS-enhanced multiuser multiple-input single-output (MU-MISO) systems. Both sum-rate maximization and power minimization problems are considered. To address their non-convexity, we develop efficient iterative algorithms to decompose them into separate several design problems, which are efficiently solved by exploiting fractional programming (FP) and Riemannian-manifold optimization techniques. Simulation results demonstrate the superiority of the proposed dual-functional active RIS architecture and the effectiveness of our proposed algorithms over various benchmark schemes.
\end{abstract}


%\vspace{-0.3 cm}
\section{Introduction}
Reconfigurable intelligent surfaces (RISs) have been proposed as a promising technology for future wireless communication systems and extensively investigated in recent years \cite{Wu}, \cite{Basar}. Specifically, an RIS is an array composed of massive passive elements. It can construct favorable wireless propagation environment between transmitters and receivers by adaptively adjusting the phase-shift of the reflected electromagnetic (EM) waves which impinge on the surface. By deploying RIS in wireless systems, the channel power gain can be effectively improved and the communication QoS can be enhanced without significant additional power consumptions.

Considering the low cost and low power consumption features, RISs have many promising applications in different wireless systems such as the multi-cell networks \cite{Ni}, multiple-input multiple-output (MIMO) communications \cite{MIMO}, \cite{lhy}, unmanned aerial vehicle (UAV) networks \cite{UAV}, and non-orthogonal multiple access (NOMA) channel \cite{NOMA} for capacity optimization, secure transmission, symbol-level precoding design, enhancing energy/spectrum efficiency \cite{Hu}-\cite{N2}, etc.
{Besides, many technologies have been investigated for RIS-based communications, e.g., deep reinforcement learning (DRL) \cite{DRL}, \cite{DRL2} and compressive sensing \cite{cs}.
An RIS architecture to achieve amplitude-and-phase-varying modulation was proposed in \cite{Tang} and the free-space path loss models for RIS-assisted wireless communications was developed in \cite{Tang2} by studying the physics and electromagnetic nature of RISs.}
Furthermore, by embedding radio frequency (RF) chains and signal processing units in the surface, RIS can work as a transmitter or receiver \cite{rf}-\cite{liur}, which can realize a low-complexity and energy-efficient virtual multi-input multi-output (MIMO) system with only several RF chains.

However, due to ``double fading'' effect (i.e., the total path loss of the cascaded transmitter-RIS-receiver link is the product of the path losses of the transmitter-RIS link and RIS-receiver link, which is usually an order-of-magnitude larger than that of the direct link), RIS should be deployed close to either the base station (BS) or users to effectively improve communication performance \cite{Active}-\cite{N2}. Besides, recent studies showed that RISs can be easily outperformed by conventional full-duplex (FD) amplify-and-forward (AF) relay unless very large RISs are employed \cite{Bjornson}, \cite{Ntontin}. In \cite{Bjornson}, the authors compared the RIS with classic decode-and-forward (DF) relaying and discovered that very large meta-surfaces are needed to surpass DF relaying.
The authors in \cite{Ntontin} carried out a performance comparison in terms of achievable rate and energy efficiency. Numerical results showed that the RIS-aided system outperforms the relay-aided network only for adequately large RIS. However, massive reflecting elements will lead to high training overhead for channel estimation \cite{Wang}, \cite{Cheng} and high power consumption to maintain the circuit operations. Consequently, the control of RIS will also become more difficult \cite{You}. Therefore, the purely passive RIS might have limited application scenarios in practical wireless systems.

In order to overcome the double fading effect in RIS-assisted systems, some proposals have been raised and attracted a lot of interest.
In \cite{relay}, the authors proposed a novel relay-aided RIS architecture consisting of two RISs connected via a full-duplex relay. This architecture can achieve the same performance with conventional RIS while only requiring much fewer reconfigurable elements. However, they only considered the single-input single-output (SISO) system and derived the theoretical upper bound of the achievable rate.
The designs of BS transmit beamforming and RIS reflection have not been well investigated. A further question is that the first RIS in their proposed system only reflects the signal towards relay and cannot simultaneously serve the users around it. In \cite{Hybrid}, the authors proposed a semi-active RIS-aided architecture, in which the RIS not only passively reflects the signal, but also actively amplifies it at the same time. However, the practical implementation of this hybrid-RIS remains an open problem, while it needs expensive RF chains.

Active RIS has been recently proposed \cite{Active}, \cite{liang} to overcome the aforementioned practical issues of passive RIS by amplifying the reflected signal with low-cost hardware.
Unlike the conventional AF relay that requires power-hungry RF chains, the active RIS
directly reflects signals in an FD manner with low-power reflection-type amplifiers. It was shown that the active-RIS aided system is able to achieve higher rate than the passive RIS owing to the amplification gain at the RIS \cite{Active}, \cite{liang}. Moreover, the authors in \cite{d_active} considered the downlink/uplink communications separately and optimized the RIS placement for rate maximization with an
active or passive RIS. Simulation results showed that the active RIS should be deployed closer to the receiver with the less amplification power of the active RIS, while the passive RIS should be deployed close either the transmitter or receiver. However, the users at the back side of RIS still cannot be served by proposed active RIS.
%
%
%%Relay can increase the energy of received signals, nevertheless, it will increases the interference and can not focus the signal to intended receiver. For the purpose of combining these benefits and overcoming these challenges,
The reflective property of RIS restricts the service coverage to only one side of the surface and the users locating behind RIS cannot be effectively served. To resolve this limitation, the novel concept of simultaneously transmitting and reflecting RIS (STAR-RIS) was proposed in \cite{STAR1}-\cite{STAR3}. In particular, the wireless signal impinging upon an element of a STAR-RIS is divided into two parts. One part is reflected to the users in front of the RIS, and the other is transmitted to the users behind the RIS. Thus, users at both sides of STAR-RIS can be served. Despite the above advantage on $360^{\circ}$ coverage, performance improvement of passive STAR-RIS is still limited due to severe signal attenuation.

To overcome the double fading effect and enhance the coverage, in this paper we propose a novel reflection and transmission dual-functional active RIS architecture.
%, which can simultaneously realize active amplification and reflection/transmission functionalities to extend the coverage and improve the QoS of all users.
Particularly, the active RIS firstly amplifies the incident signals, divides them by power split circuits, and emits them with two corresponding controllable phase-shifts towards front and back to serve all users around. Based on the proposed dual-functional active RIS architecture, we consider the joint BS beamforming and RIS reflection/transmission design in MU-MISO systems. The main contributions in this article are summarized as follows:

\begin{itemize}

\item
    We propose a novel reflection and transmission dual-functional active RIS architecture, which can simultaneously realize active amplification and reflection/transmission functionalities to extend the coverage and improve the QoS of each user. It should be noted that the proposed dual-functional active RIS is completely different with the full-duplex relays since no RF chains and complex signal processing circuits are needed. Moreover, owing to the low cost and low power consumption, RIS can be manufactured to a large scale, e.g., hundreds of elements, which can offer significant beamforming gains.
    %Moreover, for the full-duplex relay, it is inevitable to suffer from severe self-interference, while the proposed dual-functional RIS is free from that.

\item We first focus on the sum-rate maximization problem, which attempts to maximize the achievable sum-rate of all users with given BS and active RIS power consumption budgets. In order to solve this joint design problem, an efficient iterative algorithm is proposed to decompose the problem into separable designs, where the FP theory and Riemannian-manifold optimization are exploited.

\item Then, considering the additional power consumption introduced by the active RIS, we consider the power minimization problem, which aims to minimize the total power consumption of both BS and RIS while guaranteeing a certain QoS among users. An efficient iterative algorithm is proposed to optimize BS and RIS designs using the log-sum-exp approach and Riemannian-manifold algorithm after some transformations.

\item Finally, we provide extensive simulation results to demonstrate the advancement of the proposed reflection and transmission dual-functional active RIS architecture and the effectiveness of developed algorithms. In particular, we show that applying the dual-functional active RIS brings remarkable performance improvement in terms of sum-rate and power-savings. Moreover, it can be noted that the reflective active RIS and STAR-RIS design is a special case of the dual-functional active RIS we proposed. Hence, the algorithms we proposed provide unifying solution for reflective active RIS and STAR-RIS.

\end{itemize}

%The rest of this paper is organized as follows. Section \ref{s:system} introduces the concept of dual-functional active RIS and system model. Then, the sum-rate maximization and power minimization problems are investigated in Sections \ref{s:sumrate} and \ref{s:power}, respectively. Simulation results are presented in Section \ref{s:simulation}, and finally conclusions are provided in Section \ref{s:conclusion}.

\begin{figure}[t]
\centering
\includegraphics[height=2.0 in]{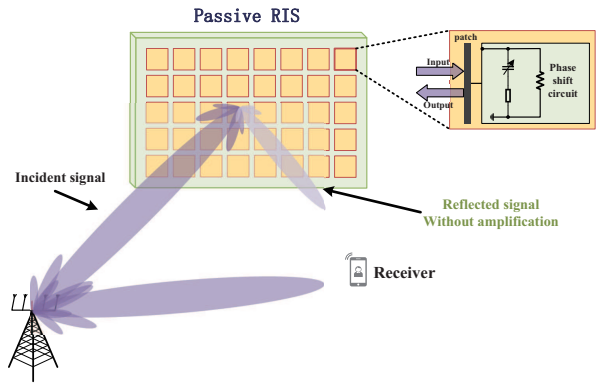}
\caption{A traditional passive RIS-aided communication system.}\label{fig:passive_RIS}
\vspace{-0.4 cm}
\end{figure}

\textit{Notations}: The following notations are used throughout this paper. $a$ is a scalar, $\mathbf{a}$ is a vector, and $\mathbf{A}$ is a matrix. $\mathbf{A}^{*}$, $\mathbf{A}^{T}$, $\mathbf{A}^{H}$, $\mathbf{A}^{-1}$ denote the conjugate, transpose, Hermitian (conjugate transpose) and inversion of $\mathbf{A}$, respectively. $\Re(\cdot)$ and $|\cdot|$ denote the real part and modulus of a complex number, respectively. diag($\mathbf{a}$) is a diagonal matrix with the entries of $\mathbf{a}$ on its main diagonal. Notation $v\sim\mathcal{CN}(0,\sigma^2)$ means that random variable $v$ is complex circularly symmetric Gaussian with zero mean and variance $\sigma^2$.  $\mathbb{C}^{M\times N}$ denotes the set of all $M\times N$ complex-valued matrices. $\mathbf{a}_m$ denotes the $m$-th element of vector $\mathbf{a}$ and $\mathbf{A}(m,n)$ denotes the $(m,n)$-th element of matrix $\mathbf{A}$.

\vspace{-0.00 cm}
\section{Dual-Functional Active RIS And RIS System Model}
\label{s:system}
\subsection{Existing Passive RIS}

\begin{figure*}[t]
\centering
\includegraphics[height=1.9 in]{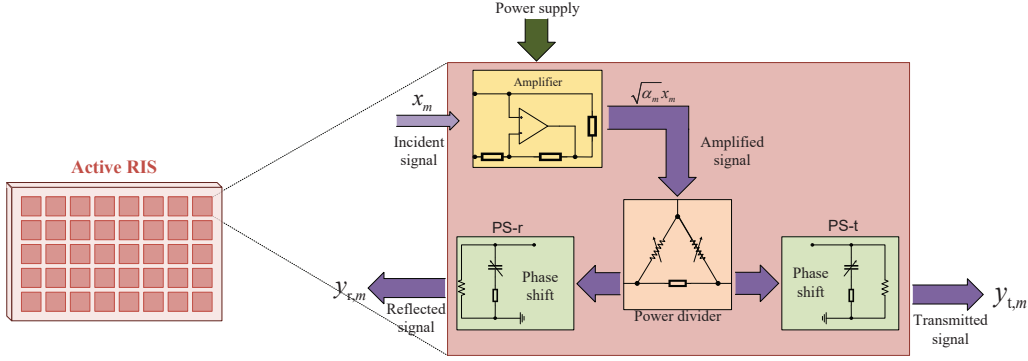}
\caption{The circuit diagram of the proposed reflection and transmission dual-functional active RIS.}\label{fig:circuit}
\vspace{-0.1 cm}
\end{figure*}

The RIS widely studied in existing works \cite{Wu}-\cite{liur} is supposed to be passive or nearly-passive, as shown in Fig \ref{fig:passive_RIS}. Passive RIS can reflect signals to achieve a signal-to-noise ratio (SNR) that grows as $M^2$, where $M$ is the number of RIS elements, while the SNR grows linearly with $M$ when using massive MIMO and half-duplex relays \cite{N2}. Besides, a more aggressive power scaling law can be formulated where the transmit power is reduced as $1/M^2$ \cite{Basar}, \cite{qing}, \cite{N2}, \cite{Kong}.

However, this expected high SNR gain can be hardly achieved in realistic communication scenarios, where the direct link between the transmitter and the receiver is usually much stronger than the reflection link, due to the significant ``double fading'' effect \cite{Active}. Therefore, thousands or even millions of RIS elements are required to compensate for this extremely severe path loss, which will result in high overhead for channel estimation and high hardware complexity \cite{Hu}, \cite{chann}.
%, \cite{Huang}

On the other hand, since passive RISs can only reflect incident wireless signals, transmitter and receiver have to be located on the same side of the RIS, as shown in Fig \ref{fig:passive_RIS}. This property limits practical deployment and gravely restricts the flexibility and effectiveness of RIS, since generally users may be located at both sides of an RIS. To tackle this issue, STAR-RIS, which is capable of simultaneously reflecting and diffracting the received signals, is proposed \cite{STAR1}, \cite{STAR2}, \cite{IOS}. However, due to the severe attenuation caused by the double-fading, the effective ``coverage'' range of the STAR-RIS is still far from being satisfactory.

\vspace{-0.0 cm}
\subsection{Proposed Dual-Functional Active RIS}

%circuit_0725

Different from conventional passive RIS which just redirects signals to the receivers in the immediate vicinity of the RIS within the same half-space bounded by the reflecting surface, the proposed dual-functional active RIS is capable of simultaneously amplifying, reflecting and transmitting the impinging electromagnetic waves towards any direction in the space.
Specifically, as shown in Fig. \ref{fig:circuit}, the signal that impinges upon each RIS element is firstly magnified by an integrated amplifier, divided by power split and fed to two controllable phase shifters (PSs), which are respectively labelled as PS-r and PS-t in Fig. \ref{fig:circuit}. Then the signals are emitted from the front side and back side of the RIS and denoted as reflected signals and transmitted signals, respectively. Thus, the proposed dual-functional active RIS not only enhances signal strength, but also extends coverage to full-space, overcoming the shortcomings of passive RIS and STAR-RIS.

Although each element of the proposed active RIS is supported by a set of additionally integrated active amplifier which needs additional power consumption to support the active load, it is noted that the power consumption of reflection-type amplifier has been decreased to the microwatt level in recent years \cite{Amato}.
This integrated amplifier can significantly improve both the energy and hardware efficiency as opposed to the traditional antenna-array type transmitter built on power-hungry and expensive RF chain components. In practice, it can be realized by many existing active components, such as the tunnel diodes \cite{Amato}, the current-inverting converter \cite{Ultrathin}, the asymmetric current mirror \cite{Bousquet}, or even some integrated chips \cite{Kishor}, which can amplify the incident signal without requiring significant energy consumption. Power split can be implemented by the power system's interharmonic components \cite{split}.
In addition, it is assumed that the amplitude and phase-shift of each RIS element can be tuned independently since they are adjusted by separate components. Overall, an active dual-functional RIS having $M$ reconfigurable elements is composed of $M$ amplifiers, $M$ power split components and $2M$ phase shifters.

%\vspace{-0.3 cm}
\subsection{Signal Model}

With the previous analysis, the reflected signal $y_{\mathrm{r},m}$ at the $m$-th element of the dual-functional active RIS can be modeled as follows
\begin{equation}
y_{\mathrm{r},m} = {\phi}_{\mathrm{r},m}\varsigma_m\sqrt{a_{m}}(x_m+v_m),
\end{equation}
where $x_m$ is the incident signal, $v_m$ is the introduced noise related to the input noise and thermal noise of the active RIS \cite{Bousquet}, $\sqrt{a_{m}}$ denotes the amplification gain of the $m$-th element, {$\varsigma_m \in [0,1]$ denotes the reflection amplitude}, and ${\phi}_{\mathrm{r},m}=e^{j \theta_{\mathrm{r},m}}$ is the phase-shift of the $m$-th element of PS-r.
Similarly, the transmitted signal $y_{\mathrm{t},m}$ at the $m$-th element can be modeled as
\begin{equation}
y_{\mathrm{t},m} = {\phi}_{\mathrm{t},m}\sqrt{1-\varsigma_m^2}\sqrt{a_{m}}(x_m+v_m),
%\underbrace{\mathbf{A}_ \Psi}_{}}_{\text {Desired signals }}+\underbrace{\mathbf{P \Theta v}_{\text {Dynamic noise }}}+\underbrace{\mathbf{n}_{\mathrm{s}}}_{\text {Static noise }}
\end{equation}
{where $\sqrt{1-\varsigma_m^2}$ denotes the transmission amplitude which needs to satisfy the energy conservation constraint} and ${\phi}_{\mathrm{t},m}=e^{j \theta_{\mathrm{t},m}}$ expresses the phase-shift of the $m$-th PS-t likewise.

Then, the reflected signal vector $\mathbf{y}_{\mathrm{r}}=[y_{\mathrm{r},1},\ldots,y_{\mathrm{r},M}]$ of an active RIS having $M$ elements can be modeled as
\begin{equation}
\mathbf{y}_\mathrm{r} = \bm{\Phi}_{\mathrm{r}}\mathbf{E}_{\mathrm{r}}\mathbf{A}\left(\mathbf{x}+\mathbf{v}\right),
\end{equation}
where $\mathbf{x}\in\mathbb{C}^M$ denotes the incident signal vector, $\mathbf{v}\in\mathbb{C}^M$ is the introduced noise vector, $\mathbf{v}\sim \mathcal{CN}\left(0, \sigma^{2}_v\mathbf{I}\right)$, $\mathbf{A} \triangleq \operatorname{diag}\left(\mathbf{a} \right)$ with $\mathbf{a}\triangleq[\sqrt{a_{1}}, \cdots, \sqrt{a_{M}}]^T$ is the amplification matrix of the active RIS,
{$\mathbf{E}_{\mathrm{r}}\triangleq \operatorname{diag}(\bm{\varsigma})= \operatorname{diag}\left([\varsigma_1,\ldots,\varsigma_M]^T\right)$ denotes the reflection amplitude coefficients,}
 $\bm{\Phi}_\mathrm{r} \triangleq\operatorname{diag}\left(\bm{\phi}_{\mathrm{r}}\right)$ and $\bm{\phi}_{\mathrm{r}}\triangleq \left[e^{j \theta_{\mathrm{r},1}}, \ldots, e^{j \theta_{\mathrm{r},M}}\right]^T$ denote reflection matrix and vector of PS-r, respectively.
Similarly, the transmitted signal vector $\mathbf{y}_{\mathrm{t}}$ can be modeled as
\begin{equation}
\mathbf{y}_\mathrm{t} = \bm{\Phi}_{\mathrm{t}}\mathbf{E}_{\mathrm{t}}\mathbf{A}(\mathbf{x}+\mathbf{v}),
\end{equation}
{where $\mathbf{E}_{\mathrm{t}}\triangleq \operatorname{diag}(\widetilde{\bm{\varsigma}})= \operatorname{diag}([\sqrt{1-\varsigma_1^2},\ldots,\sqrt{1-\varsigma_M^2}]^T)$ indicates transmission amplitude coefficients,} $\bm{\Phi}_{\mathrm{t}}\triangleq\operatorname{diag}\left(\bm{\phi}_{\mathrm{t}}\right)$ and $\bm{\phi}_{\mathrm{t}}\triangleq \left[e^{j \theta_{\mathrm{t},1}}, \ldots, e^{j \theta_{\mathrm{t},M}}\right]^T$ are defined in the same way to represent the phase-shift of PS-t.

\subsection{Power Consumption}

The majority of existing literatures only investigates the total power consumption of the active RIS \cite{Active}, \cite{d_active}.
However, the amplification circuit electronics have to work in their linear region where the output power increases linearly with the input power.  In this work, considering the limited power magnification capability
{due to the low cost of the amplifier}, we also consider the element-wise power constraint, which reads as
\begin{equation}
a_{m}\left(\left|x_m\right|^{2}+ \sigma_v^2\right)\leq p_{\mathrm{max},m}, ~~\forall m,
\end{equation}
where $p_{\mathrm{max},m}$ is the maximum power supply for the $m$-th element.
%In fact, the power of  noise is relatively small compared to the power of signal. Hence, we can ignore the noise component and consequently the power constraint for each element can be reduced to
%\begin{equation}
%a_{m}\left|x_m\right|^{2} \leq p_{\mathrm{max},m}, \forall m.
%\end{equation}
Moreover, with RIS amplification gain matrix $\mathbf{A}$, the following total power constraint of RIS is also considered
\begin{equation}
\left\|\mathbf{A}\mathbf{x}\right\|^{2}
+\sigma_{v}^{2}\|\mathbf{A}\|_F^{2}\leq P_\mathrm{R},\label{eq:pr}
\end{equation}
where $P_\mathrm{R}$ is the total power constraint at the active RIS ($P_{\mathrm{R}} \leq \sum_{m=1}^{M} p_{\mathrm{max},m}$ due to the thermal load of the circuit).

%\textcolor{blue}{
%Since each reconfigurable element uses its own amplifier and satisfies the element-wise power constraint, the total power constraint is not necessary.}
%Therefore, we only takes the element-wise power constraint into account.}

%\vspace{-0.3 cm}
\subsection{System Model}

\begin{figure}[t]
\centering
\includegraphics[height=1.9 in]{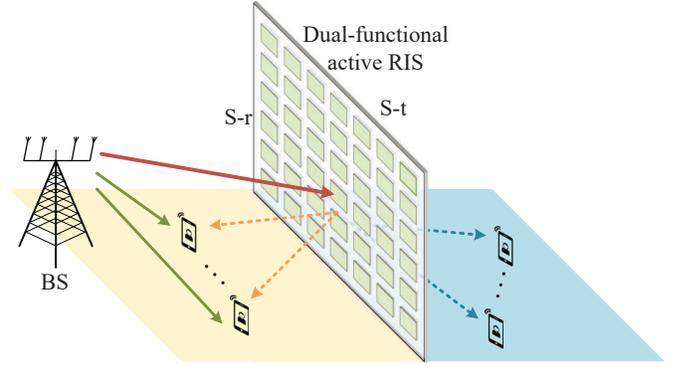}
\caption{The proposed reflection and transmission dual-functional active RIS architecture for an MU-MISO system.}\label{fig:system}
\vspace{-0.0 cm}
\end{figure}

We consider a reflection and transmission dual-functional active RIS, which is employed to assist a MU-MISO wireless communication system as illustrated in Fig. \ref{fig:system}. The BS equipped with $N$ antennas communicates with $K$ single-antenna users with the aid of a dual-functional active RIS. As shown in Fig. \ref{fig:system}, some users are located within the same half-space as BS and served by the surface facing them (which is labelled as S-r in Fig. \ref{fig:system}). Meanwhile, other users located at the opposite side are served by S-t.
%and the direct link to BS is blocked (or can be ignored due to severe path loss).
The set of all users in this system is denoted by $\mathcal{K}=\{1,2,\ldots,K\}$, while the set of users that receive the signals reflected from the S-r surface is denoted by $\mathcal{K}_\mathrm{r}$ and the set of users that receive the signals transmitted from the S-t surface
is denoted by $\mathcal{K}_\mathrm{t}$, $\mathcal{K}_\mathrm{r} \cap \mathcal{K}_\mathrm{t}=\varnothing$ and $\mathcal{K}_\mathrm{r} \cup \mathcal{K}_\mathrm{t}=\mathcal{K}$.
Different from the existing study \cite{wang}, in which the
inter-surface signal reflection is considered and the double RISs are adjusted cooperatively, in this paper, we ignore the multi-hop reflected signals due to the fact that two surfaces of the dual-functional RIS are against each other and the multi-hop reflected signals among surfaces S-r and S-t are weak. Moreover, with the rapid development of meta-surfaces, the interference between reflection and transmission can be controlled to a very low level \cite{MS1}, \cite{MS2}. Therefore, we assume that the signals emitted from S-t will not be reflected by S-r and received by the users in $\mathcal{K}_{\mathrm{r}}$ and vice versa \cite{yujsac}, \cite{myltiris}.
%\footnote{\textcolor{blue}{In future, we intend to take the mutual interference between reflection and transmission into account and investigate the cooperation between surfaces S-r and S-t.}}.
By enabling simultaneous reflection and transmission, the dual-functional active RIS extends the service coverage and enhances the strength of signals received by all users, no matter they are located at either side of the RIS.

The transmitted signal at the BS can be expressed as
\begin{equation}
\mathbf{q}= \sum_{k=1}^{K} \mathbf{w}_{k} s_{k},
\end{equation}
where $s_{k}\in \mathbb{C}, k = 1,\ldots,K$, is the transmit symbol for the $k$-th user, $\mathbb{E}\{\left|s_{k}\right|^{2}\}=1$, $\mathbf{w}_{k} \in \mathbb{C}^{N}$ is the transmit beamforming at the BS for the $k$-th user.

{Therefore, the incident signal at the active RIS can be expressed as
$\mathbf{x}=\mathbf{G}\sum_{k=1}^{K} \mathbf{w}_{k} s_{k}$, where $\mathbf{G}\in \mathbb{C}^{M\times N}$ denotes the channel from the BS to the RIS.}
The received signal at the $k$-th user, $k\in\mathcal{K}_\mathrm{r}$, which is served by the reflective S-r surface, can be modeled as
\begin{equation}
\begin{aligned}
y_{k}= &\mathbf{h}_{\mathrm{d},k}^{H}\sum_{j=1}^{K} \mathbf{w}_{j} s_{j}+\mathbf{h}_{\mathrm{r},k}^{H}\mathbf{y}_{\mathrm{r}}+n_{k}\\
=&\left(\mathbf{h}_{\mathrm{d},k}^{H}+\mathbf{h}_{\mathrm{r},k}^{H}\bm{\Phi}_{\mathrm{r}}\mathbf{E}_{\mathrm{r}}\mathbf{A} \mathbf{G}\right)\sum_{j=1}^{K} \mathbf{w}_{j} s_{j}\\
& + \mathbf{h}_{\mathrm{r},k}^{H}\bm{\Phi}_{\mathrm{r}}\mathbf{E}_{\mathrm{r}}\mathbf{A}\mathbf{v}+n_{k},~~k \in \mathcal{K}_\mathrm{r},\label{eq:y_user1}
\end{aligned}
\end{equation}
where $\mathbf{h}_{\mathrm{d},k}\in \mathbb{C}^{N}$, $\mathbf{h}_{\mathrm{r},k} \in \mathbb{C}^{M}$ denote the channels from the BS to the $k$-th user, from the RIS to the $k$-th user, respectively
\footnote{{With various advanced algorithms introduced in \cite{channel1}-\cite{channel3}
and the references therein to obtain the CSI for RIS-assisted systems, we assume perfect CSI in this paper in order to focus on exploring the potential of the reflection and transmission dual-functional active RIS-assisted system.}},
and $n_{k} \sim \mathcal{CN}\left(0, \sigma_k^{2}\right)$ denotes the complex additive white Gaussian noise with variance $\sigma_k^{2}$ at the $k$-th user. Thus, the signal-to-interference-plus-noise ratio (SINR) of the $k$-th user, $k\in\mathcal{K}_\mathrm{r}$, can be expressed as
%\begin{equation}
%\begin{aligned}
%\mathrm{SINR}_{k}= \frac{\left|\left(\mathbf{h}_{\mathrm{d},k}^{H}+\mathbf{h}_{\mathrm{r},k}^{H} \bm{\Phi}_{\mathrm{r}} \mathbf{G}\right)\mathbf{w}_{k}\right|^2}{\sum_{j \neq k}^{K+I}\left|\left(\mathbf{h}_{\mathrm{d},k}^{H}+\mathbf{h}_{\mathrm{r},k}^{H} \bm{\Phi}_{\mathrm{r}} \mathbf{G}\right)\mathbf{w}_{j} \right|^{2}+{\sigma_{k}^{2}}}.
%\end{aligned}
%\end{equation}
\begin{equation}
\begin{aligned}
\mathrm{SINR}_{k}=& \frac{\left|\widetilde{\mathbf{h}}_{k}^{H}\mathbf{w}_{k}\right|^2}{\sum_{j \neq k}^{K}\left|\widetilde{\mathbf{h}}_{k}^{H}\mathbf{w}_{j} \right|^{2}+\sigma_{v}^{2}\|\mathbf{h}_{\mathrm{r},k}^{H}\bm{\Phi}_{\mathrm{r}}\mathbf{E}_{\mathrm{r}}\mathbf{A}\|^2+{\sigma_k^{2}}}\\
\stackrel{(a)}{=}&\frac{\left|\widetilde{\mathbf{h}}_{k}^{H}\mathbf{w}_{k}\right|^2}{\sum_{j \neq k}^{K}\left|\widetilde{\mathbf{h}}_{k}^{H}\mathbf{w}_{j} \right|^{2}+\sigma_{v}^{2}\|\mathbf{h}_{\mathrm{r},k}^{H}\mathbf{E}_{\mathrm{r}}\mathbf{A}\|^2+{\sigma_k^{2}}},~k \in \mathcal{K}_\mathrm{r},
\end{aligned}
\end{equation}
wherein $(a)$ holds since $\bm{\Phi}_{\mathrm{r}}^H\bm{\Phi}_{\mathrm{r}} = \mathbf{I}$ and $\widetilde{\mathbf{h}}_{k}^{H}\triangleq \mathbf{h}_{\mathrm{d},k}^{H}+\mathbf{h}_{\mathrm{r},k}^{H} \bm{\Phi}_{\mathrm{r}}\mathbf{E}_{\mathrm{r}}\mathbf{A}\mathbf{G}$ is the equivalent channel from the BS to the $k$-th user in $\mathcal{K}_\mathrm{r}$.

%To be realistic, it is assumed that there is no line-of-sight (LoS) link between the transmitter and the users in $\mathcal{K}_\mathrm{t}$ due to the blockage of RIS and severe path attenuation.
%Based on this assumption,

Actually, there exists direct signal from the BS to users in $\mathcal{K}_\mathrm{t}$, due to scattering and reflection from all directions. The received signal at the $k$-th user in $\mathcal{K}_\mathrm{t}$ can be modeled as
\begin{equation}
\begin{aligned}
y_{k}= &\left(\mathbf{h}_{\mathrm{d},k}^{H}+\mathbf{h}_{\mathrm{r},k}^{H}\bm{\Phi}_{\mathrm{t}}\mathbf{E}_{\mathrm{t}}\mathbf{A} \mathbf{G}\right)\sum_{j=1}^{K} \mathbf{w}_{j} s_{j}\\
& + \mathbf{h}_{\mathrm{r},k}^{H}\bm{\Phi}_{\mathrm{t}}\mathbf{E}_{\mathrm{t}}\mathbf{A}\mathbf{v}+n_{k},~~k \in \mathcal{K}_\mathrm{t},\label{eq:y_user2}
\end{aligned}
\end{equation}
and the SINR can be given by
%\begin{equation}
%\begin{aligned}
%\mathrm{SINR}_{i}=\frac{|\mathbf{h}_{\mathrm{r},i}^H\boldsymbol{\Phi}_{\mathrm{t}} \mathbf{L}_\mathrm{r}\bm{\Xi}\mathbf{L}_{\mathrm{t}}^{H} \boldsymbol{\Phi}_{\mathrm{r}} \mathbf{G}\mathbf{w}_{i}|^2}
%{\sum_{j \neq i}^{K+I}\left|\mathbf{h}_{\mathrm{r},i}^H\boldsymbol{\Phi}_{\mathrm{t}} \mathbf{L}_\mathrm{r}\bm{\Xi}\mathbf{L}_{\mathrm{t}}^{H} \boldsymbol{\Phi}_{\mathrm{r}} \mathbf{G}\mathbf{w}_{j}\right|^2+\xi_i},
%\end{aligned}
%\end{equation}
\begin{equation}
\begin{aligned}
\mathrm{SINR}_{k}=\frac{|\widetilde{\mathbf{t}}_{k}^H\mathbf{w}_{k}|^2}
{\sum_{j \neq k}^{K}\left|\widetilde{\mathbf{t}}_{k}^H\mathbf{w}_{j}\right|^2
+\sigma_{v}^{2}\|\mathbf{h}_{\mathrm{r},k}^{H}\mathbf{E}_{\mathrm{t}}\mathbf{A}\|^2+\sigma_k^2},~k\in \mathcal{K}_\mathrm{t},
\end{aligned}
\end{equation}
where $\widetilde{\mathbf{t}}_{k}^{H}\triangleq \mathbf{h}_{\mathrm{d},k}^{H}+ \mathbf{h}_{\mathrm{r},k}^H\bm{\Phi}_{\mathrm{t}}\mathbf{E}_{\mathrm{t}}\mathbf{A}\mathbf{G}$ is the equivalent channel from the BS to the $k$-th user in $\mathcal{K}_\mathrm{t}$.

In this paper, we consider two typical design problems for active RIS-assisted MU-MISO systems: \textit{i}) the sum-rate maximization problem, which aims to maximize the sum-rate with given power budgets;
\textit{ii}) the power minimization problem, which minimizes the total power consumption while guaranteeing the QoS of each user. In the following sections, we will formulate and solve these two problems, respectively.

%\vspace{-0.3 cm}
\section{Sum Rate Maximization Problem}
\label{s:sumrate}

\subsection{Problem Formulation}

In this section, we aim at maximizing the sum-rate of the MU-MISO downlink system by jointly designing the transmit beamforming $\mathbf{w}_k$, $k\in\mathcal{K}$, at the BS, the amplification matrix $\mathbf{A}$ of the dual-functional active RIS, the phase-shift matrices $\boldsymbol{\Phi}_{\mathrm{r}}, \boldsymbol{\Phi}_{\mathrm{t}}$
{and the transmission and reflection amplitude coefficients $\bm{\varsigma}\triangleq [\varsigma_1,\ldots,\varsigma_M]^T$.}
%, subject to the following constraints: i) Transmit power is less than the power budget $P_\mathrm{T}$ at the BS, i.e., $\sum_{k=1}^{K}\left\|\mathbf{w}_{k}\right\|^2 \leq P_\mathrm{T}$; ii) The power constraint of each element of the active RIS i.e., $a_m\left|\mathbf{g}_m^H\sum_{k=1}^{K}\mathbf{w}_k\right|^2+a_m\sigma_v^2\leq P_m, \forall m$, where $\mathbf{g}_m$ denoting the equivalent channel from BS to the $m$-th element of RIS is the m-th row of the matrix $\mathbf{G}$; iii) Total power constraint at the dual-functional active RIS, i.e., $\sum_{k=1}^{K}\left\|\mathbf{P}\mathbf{Gw}_{k}\right\|^{2}+\sigma_{v}^{2}\|\mathbf{P}\|_F^{2}\leq P_\mathrm{R}$, ($P_\mathrm{R}\leq \sum_{m=1}^{M}P_m$ due to the thermal load of the circuit); iv) Power ratio between the reflected and transmitted signals $\varsigma_m\in[0,1], \forall m$; v) Constant amplitude of reflection coefficients, i.e. $|\bm{\Phi}_{\mathrm{r}}(m)|=1, |\bm{\phi}_{\mathrm{t}}(m)|=1, \forall m$. Therefore,
The corresponding optimization problem can be formulated as
\begin{subequations}\label{eq:problem1}
\begin{align}
\max_{\mathbf{w}_{k},\boldsymbol{\Phi}_{\mathrm{r}},\boldsymbol{\Phi}_{\mathrm{t}}, \mathbf{A}, \bm{\varsigma}} & \sum_{k=1}^{K} \log _{2}\left(1+\operatorname{SINR}_{k}\right)\label{eq:problem1q} \\
\mathrm{s.t.} ~~~~
& \sum_{k=1}^{K}\left\|\mathbf{w}_{k}\right\|^2 \leq P_\mathrm{T}, \label{eq:problem12}\\
& \sum_{k=1}^{K}\left\|\mathbf{A}\mathbf{Gw}_{k}\right\|^{2}
+\sigma_{v}^{2}\|\mathbf{A}\|_F^{2}\leq P_\mathrm{R},\label{eq:problem13}\\
&a_m(\sum_{k=1}^{K}|\mathbf{g}_m^H\mathbf{w}_k|^2+\sigma_v^2)\leq p_{\mathrm{max},m},~~ \forall m,\label{eq:problem14}\\
&\varsigma_m\in[0,1], ~~\forall m,\label{eq:problem15}\\
& |\bm{\phi}_{\mathrm{r},m}|=1, ~~|\bm{\phi}_{\mathrm{t},m}|=1,~~\forall m. \label{eq:problem16}
\end{align}
\end{subequations}
%\end{small}
where $P_\mathrm{T}$ is the power budget at the BS, $\mathbf{g}_m^H$ denotes the equivalent channel from the BS to the $m$-th element of RIS, i.e., the $m$-th row of the matrix $\mathbf{G}$.

\begin{figure*}[!t]
\normalsize
\setcounter{equation}{15}
\begin{equation} \label{eq:reR2}
g(\mathbf{w}_{k},\boldsymbol{\Phi}_{\mathrm{r}},\boldsymbol{\Phi}_{\mathrm{t}}, \mathbf{A}, \bm{\varsigma}, \bm{\gamma},\bm{\tau}) =\sum_{k\in \mathcal{K}_\mathrm{r}}
\left(2\sqrt{1+\gamma_{k}}\Re\left\{\tau_k^{\ast}\widetilde{\mathbf{h}}_{k}^H\mathbf{w}_{k}\right\}
-\left|\tau_k\right|^2\left(\sum_{j=1}^{K}\left|\widetilde{\mathbf{h}}_{k}^H\mathbf{w}_{j} \right|^{2}+\sigma_{v}^{2}\|\mathbf{h}_{\mathrm{r},k}^{H}\mathbf{E}_{\mathrm{r}}\mathbf{A}\|^2+{\sigma_k^{2}}\right)\right)
\end{equation}
\begin{equation*}
+\sum_{k\in \mathcal{K}_\mathrm{t}}
\left(2\sqrt{1+\gamma_{k}}\Re\left\{\tau_k^{\ast}\widetilde{\mathbf{t}}_{k}^H\mathbf{w}_{k}\right\}
-\left|\tau_k\right|^2\left(\sum_{j=1}^{K}\left|\widetilde{\mathbf{t}}_{k}^H\mathbf{w}_{j} \right|^{2}+\sigma_{v}^{2}\|\mathbf{h}_{\mathrm{r},k}^{H}\mathbf{E}_{\mathrm{t}}\mathbf{A}\|^2+{\sigma_k^{2}}\right)\right)
\end{equation*}
\hrulefill
\vspace*{2pt}
\end{figure*}

Obviously, the optimization problem \eqref{eq:problem1} is non-convex and difficult to solve due to the unit-modulus constraints and the coupling between variables. To tackle these difficulties, {in the following, we propose an iterative BS beamforming and RIS design algorithm based on FP theory.}

%\vspace{-0.3 cm}
%\vspace{-0.2 cm}
\subsection{FP-Based Transformation of Objective Function}

We first equivalently transform the original problem into a more tractable form based on the theory of FP. With the transformed objective function, BS beamforming, RIS reflection/transmission phase-shift matrices, amplification matrix and {amplitude coefficients} are designed iteratively.

%\subsection{FP-Based Transformation of Objective Function}
%The objective in \eqref{eq:problem1q} is a typical function of multiple fractional parameters, which is usually difficult to solve. Motivated by \cite{Shen1}, \cite{Shen2} which use FP to solve this multiple-ratio problem, we attempt to equivalently transform the original problem into a more tractable form by extracting the ratio terms SINR$_k$, $k \in\mathcal{K},$ from the logarithmic function. Based on Lagrangian dual transform \cite{Shen1}, the optimization problem \eqref{eq:problem1} is equivalent to

{The objective function in \eqref{eq:problem1q} has a typical expression of weighted sum-of-logarithmic functions of SINR$_k$, which makes the optimization problem \eqref{eq:problem1} intractable. Motivated by the closed-form FP algorithm introduced in \cite{Shen1}, \cite{Shen2}, we attempt to equivalently transform the original problem into a sum-of-ratios form by extracting the ratio terms SINR$_k$, $k \in\mathcal{K},$ from the logarithmic function.
Based on Lagrangian dual transform \cite{Shen1}, the optimization problem \eqref{eq:problem1} is equivalent to}
\vspace{-0.2cm}
\setcounter{equation}{12}
\begin{subequations}\label{eq:problem2}
\begin{align}
\max_{\mathbf{w}_{k},\boldsymbol{\Phi}_{\mathrm{r}},\boldsymbol{\Phi}_{\mathrm{t}}, \mathbf{A}, \bm{\varsigma}, \bm{\gamma}} & f(\mathbf{w}_{k},\boldsymbol{\Phi}_{\mathrm{r}},\boldsymbol{\Phi}_{\mathrm{t}}, \mathbf{A}, \bm{\varsigma},\bm{\gamma}) \label{eq:problem21}\\
\mathrm{s.t.} ~~~~~& \eqref{eq:problem12}-\eqref{eq:problem16},
\end{align}
\end{subequations}
where the objective function in \eqref{eq:problem21} is
\begin{equation} \label{eq:f1}
\begin{aligned}
f(\mathbf{w}_{k},&\boldsymbol{\Phi}_{\mathrm{r}},\boldsymbol{\Phi}_{\mathrm{t}}, \mathbf{A}, \bm{\varsigma}, \bm{\gamma})
=\sum_{k=1}^{K} \log _{2}\left(1+\gamma_{k}\right)-\sum_{k=1}^{K}\gamma_{k}\\
+&\sum_{k\in\mathcal{K}_\mathrm{r}}\frac{\left(1+\gamma_{k}\right)\left|\widetilde{\mathbf{h}}_{k}^H\mathbf{w}_{k}\right|^2}
{\sum_{j=1}^{K}\left|\widetilde{\mathbf{h}}_{k}^H\mathbf{w}_{j} \right|^{2}+\sigma_{v}^{2}\|\mathbf{h}_{\mathrm{r},k}^{H}\mathbf{E}_{\mathrm{r}}\mathbf{A}\|^2+{\sigma_k^{2}}}\\
+&\sum_{k\in\mathcal{K}_\mathrm{t}}\frac{\left(1+\gamma_{k}\right)|\widetilde{\mathbf{t}}_{k}^H\mathbf{w}_{k}|^2}
{\sum_{j=1}^{K}\left|\widetilde{\mathbf{t}}_{k}^H\mathbf{w}_{j}\right|^2+\sigma_{v}^{2}\|\mathbf{h}_{\mathrm{r},k}^{H} \mathbf{E}_{\mathrm{t}}\mathbf{A}\|^2+\sigma_k^2},
\end{aligned}
\end{equation}
with
%$\bm{\Psi}_{1}\triangleq\mathbf{A}_1\bm{\Phi}_{\mathrm{r}}=\operatorname{diag}\left(p_{1}^{1}e^{j \theta_{1}^{1}},\ldots,p_{M}^{1} e^{j \theta_{M}^{1}}\right)$ and $\bm{\Psi}_{2}\triangleq\mathbf{A}_2\bm{\phi}_{\mathrm{t}}$ denoting the compositive amplification and phase-shift response of the active RISs and
$\bm{\gamma} \triangleq [\gamma_1,\ldots,\gamma_K]^T$ being an auxiliary variable vector.
Unfortunately, optimization problem \eqref{eq:problem2} is still intractable due to the complicated form of the sum of $K$ fractional terms. Next, we apply quadratic transform \cite{Shen2} on the fractional terms to further transform them into solvable formula by introducing another auxiliary variable vector $\bm{\tau}\triangleq[\tau_1,\ldots,\tau_K]^T$. Then the optimization problem \eqref{eq:problem2} can be transformed into
\begin{subequations}
\begin{align}
\max_{\mathbf{w}_{k},\boldsymbol{\Phi}_{\mathrm{r}},\boldsymbol{\Phi}_{\mathrm{t}}, \mathbf{A}, \bm{\varsigma}, \bm{\gamma},\bm{\tau}} &  h(\bm{\gamma}) + g(\mathbf{w}_{k},\boldsymbol{\Phi}_{\mathrm{r}},\boldsymbol{\Phi}_{\mathrm{t}}, \mathbf{A}, \bm{\varsigma}, \bm{\gamma},\bm{\tau})\label{eq:problem31}\\
\mathrm{s.t.} ~~~~~&\eqref{eq:problem12}-\eqref{eq:problem16},
%\beta \leq \frac{P_\mathrm{R}}{\left|\mathbf{g}_\mathrm{t}^{H}\boldsymbol{\Phi}_{\mathrm{r}}\mathbf{G}\sum_{i = 1}^{K} \mathbf{w}_{i} \right|^{2}+\sigma_{0}^{2}}, \label{eq:problem11}\\
%& \sum_{k=1}^{K}\left\|\mathbf{w}_{k}\right\|^2 \leq P_\mathrm{T}, \label{eq:problem22}\\
%&\bm{\Phi}_{\mathrm{r}}= \mathrm{diag}\left(\bm{\Phi}_{\mathrm{r}}\right), \bm{\phi}_{\mathrm{t}}=\mathrm{diag}\left(\bm{\phi}_{\mathrm{t}}\right), \label{eq:problem13}\\
%& |\bm{\Phi}_{\mathrm{r}}(i)|=1, ~|\bm{\phi}_{\mathrm{t}}(i)|=1,~\forall i, \label{eq:problem25}\label{eq:problem3}
\end{align}
\end{subequations}
where $h(\bm{\gamma})\triangleq \sum_{k=1}^{K} \log_{2}\left(1+\gamma_{k}\right)-\sum_{k=1}^{K}\gamma_{k}$ and $g(\mathbf{w}_{k},\boldsymbol{\Phi}_{\mathrm{r}},\boldsymbol{\Phi}_{\mathrm{t}}, \mathbf{A}, \bm{\varsigma}, \bm{\gamma},\bm{\tau})$ is formulated as \eqref{eq:reR2} presented at the top of next page.

To deal with the above problem, we adopt the block coordinate ascent (BCA) methodology to alternatively update each block of variables while keeping others being fixed. The sub-problems of updating each block will be specified in details in the following.

%\vspace{-0.2 cm}
\subsection{Update Auxiliary Variable Vectors}
1) Update $\bm{\gamma}$

When other variables are fixed, the objective function $f(\mathbf{w}_{k},\boldsymbol{\Phi}_{\mathrm{r}},\boldsymbol{\Phi}_{\mathrm{t}}, \mathbf{A}, \bm{\varsigma}, \bm{\gamma})$ in (14) is a concave differentiable function with respect to the variables ${\gamma_k}$, $\forall k \in\mathcal{K}$. By setting $\frac{\partial f(\mathbf{w}_{k},\boldsymbol{\Phi}_{\mathrm{r}},\boldsymbol{\Phi}_{\mathrm{t}}, \mathbf{A}, \bm{\varsigma}, \bm{\gamma})}{\partial \gamma_{k}}$ to zero, the optimal ${\gamma}_k^{\star}$ can be found in a closed-form as

%Given $\mathbf{w}_{k}, \boldsymbol{\Phi}_{\mathrm{r}}, \boldsymbol{\Phi}_{\mathrm{t}},\mathbf{A}$ $\bm{\varsigma}$, by solving $\frac{\partial f(\mathbf{w}_{k},\boldsymbol{\Phi}_{\mathrm{r}},\boldsymbol{\Phi}_{\mathrm{t}}, \mathbf{A}, \bm{\varsigma}, \bm{\gamma})}{\partial \gamma_{k}}=0$ for \eqref{eq:f1},

\setcounter{equation}{16}
\begin{equation} \label{eq:rrr}
\gamma_{k}^{\star}=\left\{
\begin{aligned}
&\frac{|\widetilde{\mathbf{h}}_{k}^{H} \mathbf{w}_{k}|^2}{\sum_{j\neq k}^{K}\left|\widetilde{\mathbf{h}}_{k}^H\mathbf{w}_{j} \right|^{2}+\sigma_{v}^{2}\|\mathbf{h}_{\mathrm{r},k}^{H} \mathbf{E}_{\mathrm{r}}\mathbf{A}\|^2+{\sigma_k^{2}}},~k\in\mathcal{K}_\mathrm{r},\\
&\frac{|\widetilde{\mathbf{t}}_{k}^H\mathbf{w}_{k}|^2}{\sum_{j \neq k}^{K}\left|\widetilde{\mathbf{t}}_{k}^H\mathbf{w}_{j} \right|^{2}+\sigma_{v}^{2}\|\mathbf{h}_{\mathrm{r},k}^{H} \mathbf{E}_{\mathrm{t}}\mathbf{A}\|^2+{\sigma_k^{2}}},~k\in\mathcal{K}_\mathrm{t}.
\end{aligned}
\right.
\end{equation}
%whose proof is omitted due to space limitations.

2) Update $\bm{\tau}$

Similarly, given other variables,
$g(\mathbf{w}_{k},\boldsymbol{\Phi}_{\mathrm{r}},\boldsymbol{\Phi}_{\mathrm{t}}, \mathbf{A}, \bm{\varsigma}, \bm{\gamma},\bm{\tau})$ in \eqref{eq:reR2} is a concave differentiable function with respect to $\tau_k, \forall k \in\mathcal{K}$.
The optimal variable ${\tau}_k^{\star}$ can be obtained by setting $\frac{\partial g(\mathbf{w}_{k},\boldsymbol{\Phi}_{\mathrm{r}},\boldsymbol{\Phi}_{\mathrm{t}}, \mathbf{A}, \bm{\varsigma}, \bm{\gamma},\bm{\tau})}{\partial \tau_{k}}$ to zero and is expressed as

\begin{equation} \label{eq:ttt}
\tau_{k}^{\star}=\left\{
\begin{aligned}
&\frac{\sqrt{1+\gamma_{k}}\widetilde{\mathbf{h}}_{k}^{H}\mathbf{w}_{k}}
{\sum_{j=1}^{K}\left|\widetilde{\mathbf{h}}_{k}^H\mathbf{w}_{j} \right|^{2}+\sigma_{v}^{2}\|\mathbf{h}_{\mathrm{r},k}^{H} \mathbf{E}_{\mathrm{r}}\mathbf{A}\|^2+{\sigma_k^{2}}},~k\in\mathcal{K}_\mathrm{r},\\
&\frac{\sqrt{1+\gamma_{k}}\widetilde{\mathbf{t}}_{k}^{H}\mathbf{w}_{k}}
{\sum_{j=1}^{K}\left|\widetilde{\mathbf{t}}_{k}^H\mathbf{w}_{j} \right|^{2}+\sigma_{v}^{2}\|\mathbf{h}_{\mathrm{r},k}^{H} \mathbf{E}_{\mathrm{t}}\mathbf{A}\|^2+{\sigma_k^{2}}},~k\in\mathcal{K}_\mathrm{t}.
\end{aligned}
\right.
\end{equation}

%\vspace{-0.7 cm}
\subsection{Update BS Beamforming $\mathbf{w}_{k}$}
With other variables being fixed, the update of beamforming $\mathbf{w}_{k}$ can be expressed by
\begin{subequations}\label{eq:BS}
\begin{align}
\max_{\mathbf{w}_{k}}~& g(\mathbf{w}_{k},\boldsymbol{\Phi}_{\mathrm{r}},\boldsymbol{\Phi}_{\mathrm{t}}, \mathbf{A}, \bm{\varsigma}, \bm{\gamma},\bm{\tau}) \\
\mathrm{s.t.}~~ & \sum_{k=1}^{K}\left\|\mathbf{w}_{k}\right\|^2 \leq P_\mathrm{T},\\
& \sum_{k=1}^{K}\left\|\mathbf{A}\mathbf{Gw}_{k}\right\|^{2}
+\sigma_{v}^{2}\|\mathbf{A}\|_F^{2}\leq P_\mathrm{R}, \label{eq: aaa}\\
&a_m(\sum_{k=1}^{K}|\mathbf{g}_m^H\mathbf{w}_k|^2+\sigma_v^2)\leq p_{\mathrm{max},m},~~ \forall m.  \label{eq: bbb}
\end{align}
\end{subequations}
Problem \eqref{eq:BS} is a convex second-order cone programming (SOCP) \cite{socp} problem whose optimal solution can be efficiently obtained by various existing algorithms or optimization tools, e.g., CVX \cite{cvx}.

\vspace{-0.1 cm}
\subsection{Update Amplification Matrix $\mathbf{A}$}

Given $\bm{\gamma}$, $\bm{\tau}, \bm{\varsigma}$, BS beamforming $\mathbf{w}_k, \forall k$, and RIS phase-shift matrices $\bm{\Phi}_{\mathrm{r}}$, $\bm{\Phi}_{\mathrm{t}}$, the sub-problem with respect to the amplification gain $\mathbf{A}$ can be presented as
\begin{subequations}\label{p:P}
\begin{align}
\max_{\mathbf{A}}~& g(\mathbf{w}_{k},\boldsymbol{\Phi}_{\mathrm{r}},\boldsymbol{\Phi}_{\mathrm{t}}, \mathbf{A}, \bm{\varsigma}, \bm{\gamma},\bm{\tau}) \\
\mathrm{s.t.}~~
& \sum_{k=1}^{K}\left\|\mathbf{A}\mathbf{Gw}_{k}\right\|^{2}
+\sigma_{v}^{2}\|\mathbf{A}\|_F^{2}\leq P_\mathrm{R},\label{eq:p2}\\
&a_m\leq c_m,~~ \forall m,\label{eq:p3}
\end{align}
\end{subequations}
wherein $c_m\triangleq\frac{P_{\mathrm{max},m}}{\sum_{k=1}^{K}\left|\mathbf{g}_m^H\mathbf{w}_k\right|^2+\sigma_v^2}$ represents the maximum power limit of the $m$-th element of RIS. Problem \eqref{p:P} is a convex SOCP problem whose optimal solution can be efficiently obtained by various existing algorithms or optimization tools, e.g., CVX \cite{cvx}.

%\vspace{-0.3 cm}
\subsection{Update Phase-Shift Matrices $\bm{\Phi}_{\mathrm{r}}$, $\bm{\Phi}_{\mathrm{t}}$}
Given $\bm{\gamma}$, $\bm{\tau}$, amplification gain $\mathbf{A}$, {amplitude coefficients $\bm{\varsigma}$} and BS beamforming $\mathbf{w}_k, \forall k$, the sub-problem with respect to the RIS phase-shift matrices $\bm{\Phi}_{\mathrm{r}}$ and $\bm{\Phi}_{\mathrm{t}}$ can be presented as:
\begin{subequations}  \label{p:Psi}
\begin{align}
\max_{\bm{\Phi}_{\mathrm{r}},\bm{\Phi}_{\mathrm{t}}}  ~~& g(\mathbf{w}_{k},\boldsymbol{\Phi}_{\mathrm{r}},\boldsymbol{\Phi}_{\mathrm{t}}, \mathbf{A}, \bm{\varsigma}, \bm{\gamma},\bm{\tau}) \\
\mathrm{s.t.} ~~~
& |\bm{\phi}_{\mathrm{r},m}|=1, ~~|\bm{\phi}_{\mathrm{t},m}|=1,~~\forall m.
\end{align}
\end{subequations}

%We define $\bm{\phi}\triangleq [\bm{\Phi}_{\mathrm{r}};\bm{\phi}_{\mathrm{t}}]$, $\mathbf{H}_1\triangleq \left[\mathrm{diag}\left(\mathbf{h}_{k}^{H}\right); \mathbf{0}\right]$, $\mathbf{H}_2\triangleq \left[ \mathbf{0}; \mathrm{diag}\left(\mathbf{h}_{k}^{H}\right)\right]$, and $\mathbf{G}_{\mathrm{r},1}\triangleq \left[ \mathrm{diag}\left(\mathbf{g}_\mathrm{t}^{H}; \mathbf{0}\right)\right]$ and the objective function in \eqref{p:Psi} can be rearranged as
%\begin{equation}
%\begin{aligned}
%\delta = \sum_{k=1}^{K-1}\left(2\sqrt{1+r_{k}}\Re\left\{y_k^{\star}\bm{\phi}^{H}\mathbf{H}_1 \mathbf{G}\mathbf{w}_{k}\right\}-\left|y_k\right|^2\sum_{i=1}^{K}\left|\bm{\phi}^{H} \mathbf{H}_1 \mathbf{G}\mathbf{w}_{i} \right|^{2}\right)\\
%+
%2\sqrt{\beta\left(1+r_{K}\right)}\Re\left\{y_K^{\star}\bm{\phi}^{H}\mathbf{H}_2 \mathbf{g}_\mathrm{r} \bm{\phi}^{H}\mathbf{G}_{\mathrm{t},1} \mathbf{G}\mathbf{w}_{K}\right\}\\
%-\left|y_K\right|^2\beta\sum_{i=1}^{K}\left|\bm{\phi}^{H}\mathbf{H}_2 \mathbf{g}_\mathrm{r}\bm{\phi}^{H}\mathbf{G}_{\mathrm{r},1}\mathbf{G}\mathbf{w}_{i}\right|^2
%\end{aligned}
%\end{equation}
Considering that the RIS phase-shifts $\bm{\Phi}_\mathrm{r} \triangleq\operatorname{diag}\left(\bm{\phi}_{\mathrm{r}}\right)$ and $\bm{\Phi}_\mathrm{t} \triangleq\operatorname{diag}\left(\bm{\phi}_{\mathrm{t}}\right)$ are irrelevant, we turn to separately optimize either of them. By defining
\begin{subequations}
\begin{align}
\mathbf{r}_{k,j}&\triangleq
\left[\mathbf{h}_{\mathrm{r},k}^H\mathbf{E}_{\mathrm{r}}\mathbf{A} \mathrm{diag}\left(\mathbf{G}\mathbf{w}_{j}\right)\right]^H,
~j\in \mathcal{K},~k\in \mathcal{K}_\mathrm{r},\\
\mathbf{B}_\mathrm{r}&\triangleq \sum_{k\in \mathcal{K}_\mathrm{r}}\sum_{j=1}^{K}|\tau_k|^2 \mathbf{r}_{k,j}\mathbf{r}_{k,j}^H,\\
\mathbf{c}_\mathrm{r}&\triangleq
\sum_{k\in \mathcal{K}_\mathrm{r}}\sqrt{1+\gamma_{k}}\tau_k\mathbf{r}_{k,k}-\sum_{k\in \mathcal{K}_\mathrm{r}}\sum_{j=1}^{K}|\tau_k|^2\mathbf{r}_{k,j}\mathbf{h}_{\mathrm{d},k}^H\mathbf{w}_{j},
\end{align}
\end{subequations}
and removing constants, the optimization of $\boldsymbol{\phi}_{\mathrm{r}}$ \eqref{p:Psi} is equivalent to solving the following problem
\begin{subequations}  \label{p:Psi1}
\begin{align}
\min_{\boldsymbol{\phi}_{\mathrm{r}}}~~
&\sum_{k\in \mathcal{K}_\mathrm{r}}\left|\tau_k\right|^2\sum_{j=1}^{K}\left|\mathbf{h}_{\mathrm{d},k}^H\mathbf{w}_{j}+\mathbf{r}_{k,j}^H\boldsymbol{\phi}_{\mathrm{r}}\right|^{2}\nonumber \\
&-\sum_{k\in \mathcal{K}_\mathrm{r}}2\sqrt{1+\gamma_{k}}\Re\left\{\tau_k^{\ast}\mathbf{r}_{k,k}^H\boldsymbol{\phi}_{\mathrm{r}}\right\} \nonumber \\
&=\boldsymbol{\phi}_{\mathrm{r}}^H\mathbf{B}_\mathrm{r}\boldsymbol{\phi}_{\mathrm{r}}-2\Re\left\{\boldsymbol{\phi}_{\mathrm{r}}^H\mathbf{c}_\mathrm{r}\right\}+\bar{a}
\triangleq f(\boldsymbol{\phi}_{\mathrm{r}})\label{p:Psi1p}\\
\mathrm{s.t.}~~&|\bm{\phi}_{\mathrm{r},m}|=1,~~\forall m, \label{eq:unitpsi1}
\end{align}
\end{subequations}
where $\bar{a} \triangleq \sum_{k\in \mathcal{K}_\mathrm{r}}\sum_{j=1}^{K}\left|\tau_k\right|^2\mathbf{w}_{j}^H\mathbf{h}_{\mathrm{d},k}\mathbf{h}_{\mathrm{d},k}^H\mathbf{w}_{j}$ is a constant independent of $\boldsymbol{\phi}_{\mathrm{r}}$. It can be observed that, although the objective function in \eqref{p:Psi1p} is continuous and convex, problem \eqref{p:Psi1} is still difficult to solve due to the constant modulus constraint \eqref{eq:unitpsi1}. There are two popular methods for handling this type of constraint: non-convex relaxation and alternating minimization. However, the non-convex relaxation method always suffers a performance loss and the alternating minimization method usually has slow convergence as the number of variables increases. To effectively solve this problem, we adopt the Riemannian-manifold method \cite{CCM} for directly solving problem \eqref{p:Psi1} with very fast convergence \cite{manifolds}.
%\textcolor{blue}{
%On a manifold, each point has a neighborhood homeomorphic to Euclidean space, and the directions in which the point can move are its tangent vectors, which compose the tangent space.
%The optimization over a manifold is locally analogous to that in the Euclidean space. Therefore, optimization
%techniques that were developed for the Euclidean space, e.g., conjugate gradient (CG) and trust-region methods, are suitable on the Riemannian manifold with several operations.}
The search space in problem \eqref{p:Psi1} can be regarded as the product of $M$ complex circles, which can be given by
\begin{equation}
\mathcal{C}^{M} \triangleq\left\{\mathbf{x} \in \mathbb{C}^{M}:\left|{x}_l\right|=1, l=1,2, \cdots, M\right\},\label{eq:C}
\end{equation}
where ${x}_l$ is the $l$-th element of vector $\mathbf{x}$.

The main idea of the Riemannian-manifold algorithm is to derive a gradient descent algorithm based on the manifold space defined in \eqref{eq:C}, which is similar to the concept of the gradient descent technique developed for the conventional optimization over the Euclidean space. The most common search direction for a minimization problem is to move towards the direction opposite to the Euclidean gradient of it, which is given by
\begin{equation} \label{eq:g1}
\mathbf{g}^{(t)}=-\nabla f\left(\bm{\phi}_{\mathrm{r}}^{(t)}\right)=-2\mathbf{B}_\mathrm{r}\bm{\phi}_{\mathrm{r}}^{(t)}+2 \mathbf{c}_\mathrm{r}.
\end{equation}
where $\bm{\phi}_{\mathrm{r}}^{(t)}$ is the iteration point in the $t$-th update. Since we optimize over the manifold space, we have to find the Riemannian gradient.
%The Riemannian gradient at the current point $\bm{\phi}_{\mathrm{r}}^{(t)}\in \mathcal{C}^{M}$ is in the tangent space $\mathcal{T}_{\bm{\Phi}_{\mathrm{r}}^{(t)}}$.
Specifically, the Riemannian gradient at $\bm{\phi}_{\mathrm{r}}^{(t)}$ can be obtained by projecting the search direction $\mathbf{g}^{(t)}$ in the Euclidean space $\mathcal{T}_{\bm{\phi}_{\mathrm{r}}^{(t)}}$ using the projection operator, which can be calculated as follows
\begin{equation} \label{eq:rg1}
\mathbf{P}_{\mathcal{T}_{\bm{\phi}_{\mathrm{r}}^{(t)}}}=\mathbf{g}^{(t)}-\operatorname{Re}\left\{\mathbf{g}^{(t)}\odot \bm{\phi}_{\mathrm{r}}^{(t)}\right\} \odot \bm{\phi}_{\mathrm{r}}^{(t)}.
\end{equation}
where $\odot$ represents the Hadamard product. Then, we update the current point $\bm{\phi}_{\mathrm{r}}^{(t)}$ on the tangent space %$\mathcal{T}_{\bm{\phi}_{\mathrm{r}}^{(t)}}$
\begin{equation} \label{eq:pt}
\bar{\bm{\phi}_{\mathrm{r}}}^{(t)}=\bm{\phi}_{\mathrm{r}}^{(t)}+\beta \mathbf{P}_{\mathcal{T}_{\bm{\phi}_{\mathrm{r}}^{(t)}}}
\end{equation}
where $\beta$ is a constant step-size. In general, the obtained $\bar{\bm{\phi}}^{(t)}$ is not in $\mathcal{C}^{M}$, i.e. $\bar{\bm{\phi}}^{(t)} \notin \mathcal{C}^{M}$. Hence, it has to be mapped into the manifold $\mathcal{C}^{M}$ by using the retraction operator as follows
\begin{equation}
\bm{\phi}_{\mathrm{r}}^{(t+1)}=\bar{\bm{\phi}_{\mathrm{r}}}^{(t)} \odot \frac{1}{\left|\bar{\bm{\phi}_{\mathrm{r}}}^{(t)}\right|}.\label{eq:oppsi1}
\end{equation}
Now $\bm{\phi}_{\mathrm{r}}^{(t+1)}$ belongs to $\mathcal{C}^{M}$, which satisfies the unit constant modulus constraint.
The details of the Riemannian-manifold algorithm are presented in Algorithm \ref{alg:Riemannian} and {the computational complexity is approximated by $\mathcal{O}(M^{2})$ \cite{manifolds}, \cite{CG}}.

Obtaining $\boldsymbol{\phi}_{\mathrm{r}}$, we can solve $\boldsymbol{\phi}_{\mathrm{t}}$ by repeating the same procedure. Specifically, by defining
\begin{subequations}
\begin{align}
\mathbf{r}_{k,j} &\triangleq \left[\mathbf{h}_{\mathrm{r},k}^H \mathbf{E}_{\mathrm{t}}\mathbf{A}\mathrm{diag}\left(\mathbf{G}\mathbf{w}_{j}\right)\right]^H,~j\in \mathcal{K},~k\in \mathcal{K}_\mathrm{t},\\
\mathbf{B}_\mathrm{t}&\triangleq \sum_{k\in \mathcal{K}_\mathrm{t}}\sum_{j=1}^{K}|\tau_k|^2\mathbf{r}_{k,j}\mathbf{r}_{k,j}^H,\\
\mathbf{c}_\mathrm{t}&\triangleq
\sum_{k\in \mathcal{K}_\mathrm{t}}\sqrt{1+\gamma_{k}}\tau_k\mathbf{r}_{k,k}-\sum_{k\in \mathcal{K}_\mathrm{t}}\sum_{j=1}^{K}|\tau_k|^2\mathbf{r}_{k,j}\mathbf{h}_{\mathrm{d},k}^H\mathbf{w}_j,
\end{align}
\end{subequations}
and removing constants, the corresponding optimization problem of $\boldsymbol{\phi}_{\mathrm{t}}$ can be formulated as
\begin{subequations}  \label{p:Psi2t}
\begin{align}
\min_{\boldsymbol{\phi}_{\mathrm{t}}} ~
&\sum_{k\in \mathcal{K}_\mathrm{t}}|\tau_k|^2\sum_{j=1}^{K} \left|\mathbf{h}_{\mathrm{d},k}^H\mathbf{w}_{j}+\mathbf{r}_{k,j}^H \boldsymbol{\phi}_{\mathrm{t}}\right|^2 \nonumber\\
&-\sum_{k\in \mathcal{K}_\mathrm{t}}
2\sqrt{1+\gamma_{k}}\Re\left\{\tau_k^*\mathbf{r}_{k,k}^H\boldsymbol{\phi}_{\mathrm{t}}\right\}\nonumber\\
&=\bm{\phi}_{\mathrm{t}}^H\mathbf{B}_\mathrm{t}\bm{\phi}_{\mathrm{t}}-2\Re\left\{\bm{\phi}_{\mathrm{t}}^H\mathbf{c}_\mathrm{t}\right\}+\bar{b}
 \label{eq:Psi1p}\\
\mathrm{s.t.}~~ &|\bm{\phi}_{\mathrm{t},m}|=1,~~ \forall m, \label{eq:p2uint}
\end{align}
\end{subequations}
where $\bar{b} \triangleq \sum_{k\in \mathcal{K}_\mathrm{t}}\sum_{j=1}^{K}\left|\tau_k\right|^2\mathbf{w}_{j}^H\mathbf{h}_{\mathrm{d},k}\mathbf{h}_{\mathrm{d},k}^H\mathbf{w}_{j}$ is a constant independent of $\boldsymbol{\phi}_{\mathrm{t}}$. The optimal solution can be obtained following the identical argument as above. The search direction
can be given by
%Following the identical argument as before, the Euclidean gradient of problem \eqref{p:Psi2t} is given by
\begin{equation}
\mathbf{g}^{(t)}=-\nabla f\left(\bm{\phi}_{\mathrm{t}}^{(t)}\right)=-2\mathbf{B}_\mathrm{t}\bm{\phi}_{\mathrm{t}}^{(t)}+2 \mathbf{c}_\mathrm{t}.
\end{equation}
Then, the Riemannian-manifold-based algorithm in Algorithm \ref{alg:Riemannian} can be applied to obtain the phase-shift $\bm{\phi}_{\mathrm{t}}$.
Since the RIS phase-shifts $\boldsymbol{\phi}_{\mathrm{r}}$ and $\boldsymbol{\phi}_{\mathrm{t}}$ are separate, they can be optimized in parallel to reduce computational time.

\begin{algorithm}[!t]\begin{small}
\caption{Riemannian-manifold-based RIS Phase-Shift Design}
\label{alg:Riemannian}
\begin{algorithmic}[1]
\REQUIRE $f(\boldsymbol{\phi}_{\mathrm{r}}^{(1)})$, $\boldsymbol{\phi}_{\mathrm{r}}^{(1)}\in\mathcal{M}$, $T_\text{max}$, $\delta_\mathrm{th}$.
\ENSURE $\boldsymbol{\phi}_{\mathrm{r}}^\star$.
    \STATE {Initialize $t=1, \delta = \infty$.}
    \STATE{\textbf{while}} $t\leq T_\text{max}$ and $\delta\geq \delta_\mathrm{th}$ \textbf{do}
        \STATE{\hspace{0.2 cm}Calculate the Euclidean gradient $\mathbf{g}^{(t)}$ by solving (\ref{eq:g1}).}
        \STATE{\hspace{0.2 cm}Calculate the Riemannian gradient $\mathbf{P}_{\mathcal{T}_{\bm{\phi}_{\mathrm{r}}^{(t)}}}$ by solving (\ref{eq:rg1}).}
        \STATE{\hspace{0.2 cm}Update the tangent space according to (\ref{eq:pt}).}
         \STATE{\hspace{0.2 cm}Update $\bm{\phi}_{\mathrm{r}}^{(t+1)}$ by retracting $\bar{\bm{\phi}}^{(t)}$ to the complex circle manifold $\mathcal{M}$ according to (\ref{eq:oppsi1}).}
        \STATE{\hspace{0.2 cm}$\delta = \frac{\left|f(\boldsymbol{\phi}_1^{(t+1)})-f(\boldsymbol{\phi}_1^{(t)})\right|}{f(\boldsymbol{\phi}_1^{(t+1)})}$}.
        \STATE{\hspace{0.2 cm}$t = t+1$.}
    \STATE{\textbf{end while}}
    \end{algorithmic}\end{small}
\end{algorithm}

%\vspace{-0.7 cm}

{
\subsection{Update Amplitude Coefficients}}

After obtaining other variables, {the problem of solving
amplitude coefficients $\bm{\varsigma}$ can be expressed as}
\begin{subequations}\label{p:varsigma}
\begin{align}
\max_{{\bm{\varsigma}}}~& g(\mathbf{w}_{k},\boldsymbol{\Phi}_{\mathrm{r}},\boldsymbol{\Phi}_{\mathrm{t}}, \mathbf{A}, \bm{\varsigma}, \bm{\gamma},\bm{\tau}) \\
\mathrm{s.t.}~~
& \varsigma_m\in[0,1], ~~\forall m.\label{eq:v}
\end{align}
\end{subequations}
By defining
\begin{equation}
\mathbf{u}_{k,j}\triangleq
\left\{\begin{aligned}
&[\mathbf{h}_{\mathrm{r},k}^H\mathrm{diag}\left(\bm{\Phi}_{\mathrm{r}}\mathbf{A} \mathbf{G}\mathbf{w}_{j}\right)]^H,
~j\in \mathcal{K},~k\in \mathcal{K}_\mathrm{r},\\
&[\mathbf{h}_{\mathrm{r},k}^H\mathrm{diag}\left(\bm{\Phi}_{\mathrm{t}}\mathbf{A} \mathbf{G}\mathbf{w}_{j}\right)]^H,
~j\in \mathcal{K},~k\in \mathcal{K}_\mathrm{t},
\end{aligned}
\right.
\end{equation}
\begin{align}
\mathbf{Q}_\mathrm{r}&\triangleq\sum_{k\in \mathcal{K}_\mathrm{r}}\sum_{j=1}^{K}\left|\tau_k\right|^2\mathbf{u}_{k,j}\mathbf{u}_{k,j}^H\nonumber\\
&+\sum_{k\in \mathcal{K}_\mathrm{r}}\left|\tau_k\right|^2\sigma_{v}^{2}
\mathrm{diag}\left(\mathbf{h}_{\mathrm{r},k}^H\right)\mathbf{A}\mathbf{A}^H\mathrm{diag}\left(\mathbf{h}_{\mathrm{r},k}^H\right)^H,\\
\mathbf{Q}_\mathrm{t}&\triangleq\sum_{k\in \mathcal{K}_\mathrm{t}}\sum_{j=1}^{K}\left|\tau_k\right|^2\mathbf{u}_{k,j}\mathbf{u}_{k,j}^H\nonumber \\
&+\sum_{k\in\mathcal{K}_\mathrm{t}}\left|\tau_k\right|^2\sigma_{v}^{2}
\mathrm{diag}\left(\mathbf{h}_{\mathrm{r},k}^H\right)\mathbf{A}\mathbf{A}^H\mathrm{diag}\left(\mathbf{h}_{\mathrm{r},k}^H\right)^H,\\
\mathbf{b}_\mathrm{r}&\triangleq\sum_{k\in \mathcal{K}_\mathrm{r}}
\sqrt{1+\gamma_{k}}\tau_k\mathbf{u}_{k,k}-\sum_{k\in \mathcal{K}_\mathrm{r}}\sum_{j=1}^{K}\left|\tau_k\right|^2\mathbf{u}_{k,j}\mathbf{h}_{\mathrm{d},k}^{H}\mathbf{w}_{j},\\
\mathbf{b}_\mathrm{t}&\triangleq\sum_{k\in \mathcal{K}_\mathrm{t}}
\sqrt{1+\gamma_{k}}\tau_k\mathbf{u}_{k,k}-\sum_{k\in \mathcal{K}_\mathrm{t}}\sum_{j=1}^{K}\left|\tau_k\right|^2\mathbf{u}_{k,j}\mathbf{h}_{\mathrm{d},k}^{H}\mathbf{w}_{j},
%\end{gather}
%\end{eqnarray}
\end{align}
%\end{equation}
problem \eqref{p:varsigma} can be rearranged as
\begin{subequations}\label{p:revarsigma}
\begin{align}
\min_{{\bm{\varsigma}}}~& \bm{\varsigma}_{\mathrm{r}}^H\mathbf{Q}_\mathrm{r}\bm{\varsigma}_{\mathrm{r}}-2\Re\left\{\bm{\varsigma}_{\mathrm{r}}^H\mathbf{b}_\mathrm{r}\right\}
+\bm{\varsigma}_{\mathrm{t}}^H\mathbf{Q}_\mathrm{t}\bm{\varsigma}_{\mathrm{t}}-2\Re\left\{\bm{\varsigma}_{\mathrm{t}}^H\mathbf{b}_\mathrm{t}\right\}\label{eq:rek}\\
\mathrm{s.t.}~~
& \varsigma_m\in[0,1], ~~\forall m.\label{eq:rev}
\end{align}
\end{subequations}
To effectively solve this problem, we propose to iteratively design each element of the vector $\bm{\varsigma}$ until convergence. To facilitate this calculation, we first split the objective \eqref{eq:rek} as
\begin{equation}
\begin{aligned}
& \bm{\varsigma}_{\mathrm{r}}^H\mathbf{Q}_\mathrm{r}\bm{\varsigma}_{\mathrm{r}}-2\Re\left\{\bm{\varsigma}_{\mathrm{r}}^H\mathbf{b}_\mathrm{r}\right\}
+\bm{\varsigma}_{\mathrm{t}}^H\mathbf{Q}_\mathrm{t}\bm{\varsigma}_{\mathrm{t}}-2\Re\left\{\bm{\varsigma}_{\mathrm{t}}^H\mathbf{b}_\mathrm{t}\right\} \\
=& \sum_{m=1}^{M}\sum_{n=1}^{M} \mathbf{Q}_\mathrm{r}(m, n)\varsigma_m\varsigma_n-2 \Re\left\{\sum_{m=1}^{M} \varsigma_m{b}_{\mathrm{r},m}\right\}\\
&+\sum_{m=1}^{M}\sum_{n=1}^{M} \mathbf{Q}_\mathrm{t}(m, n)\sqrt{1-\varsigma_m^2}\sqrt{1-\varsigma_n^2}\\
&-2 \Re\left\{\sum_{m=1}^{M} \sqrt{1-\varsigma_m^2}b_{\mathrm{t},m}\right\},
\end{aligned}
\end{equation}
where ${b}_{\mathrm{r},m}, b_{\mathrm{t},m}$ is the $m$-th element of $\mathbf{b}_{\mathrm{r}}, \mathbf{b}_{\mathrm{t}}$, respectively. Since $\mathbf{Q}_\mathrm{r} = \mathbf{Q}_\mathrm{r}^H$ and $\mathbf{Q}_\mathrm{t} = \mathbf{Q}_\mathrm{t}^H$, the objective function with respect to the element $ {\varsigma}_m$ is given by
\begin{equation}
\begin{aligned}
&2\Re\left\{\left(\sum_{n \neq m} \mathbf{Q}_\mathrm{r}(m, n) \varsigma_n-{b}_{\mathrm{r},m}\right) \varsigma_m\right\}+\mathbf{Q}_\mathrm{r}(m, m)\varsigma_m^{2}\\
&+2\Re\left\{\left(\sum_{n \neq m} \mathbf{Q}_\mathrm{t}(m, n) \sqrt{1-\varsigma_n^2}-b_{\mathrm{t},m}\right) \sqrt{1-\varsigma_m^2}\right\} \\
&+\mathbf{Q}_\mathrm{t}(m, m)\left(1-\varsigma_m^2\right).
\end{aligned}
\end{equation}
Hence, the sub-problem with respect to $\varsigma_m$ while fixing other elements can be formulated as
\begin{subequations}\label{p:lvarsigma}
\begin{align}
\min_{{\varsigma_m}} ~&a_m\varsigma_m^{2} + r_m\varsigma_m + c_m\left(1-\varsigma_m^2\right) +t_m\sqrt{1-\varsigma_m^2} \\
\mathrm{s.t.} ~~&\varsigma_m\in[0,1], ~~\forall m,
\end{align}
\end{subequations}
where we define
\begin{align}
a_m &\triangleq \mathbf{Q}_\mathrm{r}(m, m),\\
c_m &\triangleq \mathbf{Q}_\mathrm{t}(m, m),\\
r_m &\triangleq 2\Re\left\{\sum_{n \neq m} \mathbf{Q}_\mathrm{r}(m, n) \varsigma_n-{b}_{\mathrm{r},m}\right\},\\
t_m &\triangleq 2\Re\left\{\sum_{n \neq m} \mathbf{Q}_\mathrm{t}(m, n) \sqrt{1-\varsigma_n^2}-b_{\mathrm{t},m}\right\}.
\end{align}

\begin{algorithm}[!t]\begin{small}
\caption{{Amplitude Coefficients Design}}
\label{alg:ratio}
\begin{algorithmic}[1]
\REQUIRE $\mathbf{Q}_\mathrm{r}$, $\mathbf{Q}_\mathrm{t}$, $\mathbf{b}_\mathrm{r}$, $\mathbf{b}_\mathrm{t}$.
\ENSURE  $\bm{\varsigma}^\star$.
    \STATE {Initialize $\bm{\varsigma}$.}
    \STATE{\textbf{while} no convergence \textbf{do}}
        \STATE{\hspace{0.4 cm}\textbf{for} $m = 1:M$ \textbf{do}.}
        \STATE{\hspace{0.6 cm}Calculate $g_m(0)$, $g_m(1-\triangle)$.}
        \STATE{\hspace{0.6 cm}if $g_m(0) \geq 0$ $\&\&$ $g_m(1-\triangle) \geq 0$}
        \STATE{\hspace{0.8 cm}$\varsigma_m^\star$ = 0.}
        \STATE{\hspace{0.6 cm}if $g_m(0) <0$ $\&\&$ $g_m(1-\triangle) < 0$}
        \STATE{\hspace{0.8 cm}$\varsigma_m^\star$ = 1.}
        \STATE{\hspace{0.6 cm}if $g_m(0) < 0$ $\&\&$ $g_m(1-\triangle) > 0$}
        \STATE{\hspace{0.8 cm}$\frac{dg_m}{d \varsigma_m}=0$.}
        \STATE{\hspace{0.6 cm}if $g_m(0) > 0$ $\&\&$ $g_m(1-\triangle) < 0$}
        \STATE{\hspace{0.8 cm}$\varsigma_m^\star = \arg \min \{g_m(0), g_m(1-\triangle)\}$ .}
        \STATE{\hspace{0.4 cm}\textbf{end for}}
    \STATE{\textbf{end while}}
    \end{algorithmic}\end{small}
\end{algorithm}

It can be noted that the objective function in problem \eqref{p:lvarsigma} is an ellipse. Hence, there is at most one stationary point between 0 and 1. By judging the derivative values at 0 and 1, the optimal solution can easily be obtained. Specifically, we firstly derive the derivative $g_m$ and calculate the value at 0 and 1
\begin{subequations}
\begin{align}
g_m &= 2(a_m-c_m)\varsigma_m + r_m -\frac{t_m\varsigma_m}{\sqrt{1-\varsigma_m^2}},\\
g_m(0) &= r_m,\\
g_m(1-\triangle) &= 2(a_m-c_m)(1-\triangle)+ r_m - \frac{t_m(1-\triangle)}{\sqrt{1-(1-\triangle)^2}},
\end{align}
\end{subequations}
where $\triangle\rightarrow 0 $ is an introduced small number in order to approach the derivative at 1. There are four possible scenarios:

\begin{enumerate}
\item $g_m(0) \geq 0$ and $g_m(1-\triangle) \geq 0$. The objective function in \eqref{p:lvarsigma} is monotonically increasing and the optimal value is 0.

\item $g_m(0) < 0$ and $g_m(1-\triangle) < 0$. The objective function in \eqref{p:lvarsigma} is monotonically decreasing and the optimal value is 1.

\item $g_m(0) < 0$ and $g_m(1-\triangle) > 0$. The objective function in \eqref{p:lvarsigma} is first decreasing and then increasing, and the optimal value can be obtained by letting $\frac{dg_m}{d \varsigma_m}=0$.

\item $g_m(0) > 0$ and $g_m(1-\triangle) < 0$. The objective function in \eqref{p:lvarsigma} is first increasing and then decreasing, and the optimal value is the point at which the objective function is allowed to take its minimum value.
\end{enumerate}
{The whole transmission and reflection amplitude coefficients optimization is summarized in Algorithm \ref{alg:ratio}. Since the amplitude coefficients design algorithm is linear}, the overall time complexity is about $\mathcal{O}\left\{I_\varsigma M\right\}$, where $I_\varsigma$ denotes the number of iterations when updating $\bm{\varsigma}$ in Algorithm 2.

%\textcolor{blue}{
%We also include two other reflection and transmission modes: \textit{i}) Space Division (SD) mode, i.e., elements are divided into two groups, one group operates for the reflection functionality while the other group operates for the transmission functionality. In this mode, the power ratio $\varsigma_m$ is fixed as 1 or 0 for these two groups, respectively. \textit{ii}) Equal Power (EP) mode, i.e., the powers of reflected signal and transmitted signal are fixed at the same value as $\varsigma_m = {1}/{\sqrt{2}}$.}

The overall procedure of solving $\bm{\varsigma}$ is summarized in Algorithm \ref{alg:ratio}. Now, the complete procedure of finding the optimal auxiliary vectors $\bm{\gamma}^{\star}$ and $\bm{\tau}^{\star}$, the BS beamforming $\mathbf{w}_k^{\star}$, amplification gain $\mathbf{A}^{\star}$, RIS reflection matrices $\bm{\Phi}_{\mathrm{r}}^{\star}, \bm{\Phi}_{\mathrm{t}}^{\star}$, and {amplitude coefficients  $\bm{\varsigma}^{\star}$} is straightforward.
The initial point of amplification gain matrix $\mathbf{A}$ is obtained by setting $\mathbf{A}=a_{\max } \mathbf{I}_{M}$ with $a_{\max }^2 =\frac{P_\mathrm{R}}{\sum_{k=1}^{K}\left\|\mathbf{Gw}_{k}\right\|^{2}
+\sigma_{v}^{2}M}$.
The powers of reflected signal and transmitted signal are fixed at the same value, {i.e., $\varsigma_m = {1}/{\sqrt{2}}, ~\forall m$.} We just simply give random values of phase-shift $\boldsymbol{\phi}_{\mathrm{r}}$ and $\boldsymbol{\phi}_{\mathrm{t}}$. The typical MMSE beamforming as the initial value of transmit beamforming $\mathbf{w}_k$ is given by
$\overline{\mathbf{w}}_{k}=(\sum_{k=1}^{K}\widetilde{\mathbf{h}}_{k}\widetilde{\mathbf{h}}_{k}^{H}+\widetilde{\sigma_k}^2 \mathbf{I})^{-1}\widetilde{\mathbf{h}}_{k}, k \in \mathcal{K}_\mathrm{r},$ with $\widetilde{\sigma_k}^2\triangleq\sigma_{v}^{2}\|\mathbf{h}_{\mathrm{r},k}^{H}\mathbf{E}_{\mathrm{r}}\mathbf{A}\|^2+{\sigma_{k}^{2}}$, and $\overline{\mathbf{w}}_{k}=(\sum_{k=1}^{K}\widetilde{\mathbf{t}}_{k}\widetilde{\mathbf{t}}_{k}^{H}+\widetilde{\sigma_k}^2 \mathbf{I})^{-1}\widetilde{\mathbf{t}}_{k}, k \in \mathcal{K}_\mathrm{t}$ with $\widetilde{\sigma_k}^2\triangleq\sigma_{v}^{2}\|\mathbf{h}_{\mathrm{r},k}^{H}\mathbf{E}_{\mathrm{t}}\mathbf{A}\|^2+{\sigma_{k}^{2}}$.
Then the transmit beamforming is further normalized to satisfy the transmit power budget by
$\mathbf{w}_{k}=\frac{P_\mathrm{T}{\overline{\mathbf{w}}_{k}}}{\sqrt{ \sum_{k=1}^{K}\left\|\overline{\mathbf{w}}_{k}\right\|_{2}^{2}}}, \forall k.$

With appropriate initialization, we iteratively update $\bm{\gamma}$, $\bm{\tau}$, $\mathbf{w}_k$, $\mathbf{A}$,
$\bm{\Phi}_{\mathrm{r}}, \bm{\Phi}_{\mathrm{t}}$  and $\bm{\varsigma}$ until convergence. For clarity, the proposed FP-based BS beamforming and RIS design algorithm is summarized in Algorithm \ref{alg:allsr}.

%\vspace{-0.5 cm}

\subsection{Complexity Analysis}
In this subsection, we provide a brief computational complexity analysis for the proposed joint BS beamforming and RIS design for the sum-rate maximization problem, i.e., Algorithm \ref{alg:allsr}.

The overall computational complexity of proposed algorithm is mainly caused by the update of variables. In each iteration, obtaining the optimal solution of $\bm{\gamma}$ and $\bm{\tau}$ requires approximately $\mathcal{O}(K^2 M^2)$ and $\mathcal{O}(K(K+1) M^2)$ operations, respectively; updating the BS transmit beamforming $\mathbf{w}_k$ requires about $\mathcal{O}((KN)^{3}(M+2)^{1.5})$ operations; the RIS amplification gain matrix $\mathbf{A}$ has a complexity of approximately $\mathcal{O}(M^{3}(M+1)^{1.5})$. {Therefore, the total computational complexity of Algorithm 3 can be approximated by $\mathcal{O}\left(I_R\left(M^{4.5}+(KN)^{3}M^{1.5}+K^2 M^2+M^{2}+I_\varsigma M\right)\right)$,} wherein $I_R$ represents the required number of iterations for the algorithm convergence.

\begin{algorithm}[!t]\begin{small}
\caption{Joint BS Beamforming and RIS Design for the Sum-Rate Maximization Problem}
\label{alg:allsr}
\begin{algorithmic}[1]
\REQUIRE $\mathbf{h}_{\mathrm{d},k}^H$, $\mathbf{h}_{\mathrm{r},k}^H$, $\mathbf{G}$, $\sigma^2$, $\sigma_v^2$, $P_\mathrm{T}$, $P_\mathrm{R}$, $P_m, \forall m$, $T_\text{max}$, $\delta_\text{th}$.
\ENSURE $\mathbf{w}_k^\star$, $\bm{\Phi}_{\mathrm{r}}^\star, \bm{\Phi}_{\mathrm{t}}^\star$, $\mathbf{A}^\star$ and $\bm{\varsigma}^\star$.
    \STATE {Initialize $\mathbf{w}_k$, $\bm{\Phi}_{\mathrm{r}}, \bm{\Phi}_{\mathrm{t}}$, $\mathbf{A}$, $\bm{\varsigma}$, $t = 1$, $\delta = \infty$, $R_{\text{temp}}=0$.}
    \STATE{\textbf{while} $t\leq T_\text{max}$ and $\delta \geq \delta_\text{th}$ \textbf{do}}
        \STATE{\hspace{0.4 cm}$R_{\text{pre}} = R_{\text{temp}}$.}
        \STATE{\hspace{0.4 cm}Update $\bm{\gamma}_k^\star, \forall k$ by (\ref{eq:rrr}).}
        \STATE{\hspace{0.4 cm}Update $\bm{\tau}_k^\star, \forall k$ by (\ref{eq:ttt}).}
        \STATE{\hspace{0.4 cm}Update BS beamforming $\mathbf{w}_k^\star, \forall k$ by (\ref{eq:BS}).}
        \STATE{\hspace{0.4 cm}Update RIS amplification gain $\mathbf{A}^\star$ by (\ref{p:P}).}
        \STATE{\hspace{0.4 cm}Update reflection matrices $\bm{\Phi}_{\mathrm{r}}^\star, \bm{\Phi}_{\mathrm{t}}^\star$ using Algorithm \ref{alg:Riemannian}.}
        \STATE{\hspace{0.4 cm}Update amplitude coefficients $\bm{\varsigma}^\star$ using Algorithm \ref{alg:ratio}.}
        \STATE{\hspace{0.4 cm}Calculate sum rate $R_{\text{temp}}=\sum_{k=1}^{K} \log_{2}\left(1+\operatorname{SINR}_{k}\right)$.}
        \STATE{\hspace{0.4 cm}$\delta = \frac{\left|R_{\text{temp}} - R_{\text{pre}}\right|}{R_{\text{temp}}}$.}
        \STATE{\hspace{0.4 cm}$t=t+1$.}
    \STATE{\textbf{end while}}
    \end{algorithmic}\end{small}
\end{algorithm}

\vspace{0.5 cm}
\section{Power Minimization Problem}
\label{s:power}
\vspace{0.1 cm}
\subsection{Problem formulation}

In this section, we aim to minimize the total power consumed by the BS and active RIS subject to the $\operatorname{SINR}$ constraint of each user. With the previous analysis, the power minimization problem can be formulated as
\begin{subequations}  \label{p:Minpower}
\begin{align}
\min_{\mathbf{w}_{k},\boldsymbol{\Phi}_{\mathrm{r}},\boldsymbol{\Phi}_{\mathrm{t}}, \mathbf{A}, \bm{\varsigma}}
&{\alpha\sum_{k= 1}^{K} \left\|\mathbf{w}_{k}\right\|^{2}+(1-\alpha)\sum_{k= 1}^{K}\left\|\mathbf{A}\mathbf{G}\mathbf{w}_{k}\right\|^{2} \label{eq:Minpowerp}}\\
\mathrm{s.t.}~~~~ &\operatorname{SINR}_{k} \geq \gamma_k, ~~k\in\mathcal{K},\label{eq:sinr}\\
&a_m(\sum_{k=1}^{K}|\mathbf{g}_m^H\mathbf{w}_k|^2+\sigma_v^2)\leq p_{\mathrm{max},m}, ~~\forall m,\\
&\varsigma_m\in[0,1], ~~\forall m,\label{eq:ratio}\\
&|\bm{\phi}_{\mathrm{r},m}|=1,|\bm{\phi}_{\mathrm{t},m}|=1,~~\forall m, \label{eq:unit}
\end{align}
\end{subequations}
{where $\alpha\in(0,1)$ acts as the weighting factor }and $\gamma_k > 0 $ is the minimum SINR requirement of user $k$. Although the objective function of problem \eqref{p:Minpower} is convex, it is challenging to solve due to the non-convex constraint in \eqref{eq:sinr} where the variables are coupled. In the following, to tackle this difficulty, we propose to decompose this multivariate problem into several sub-problems and solve each of them iteratively.

Prior to solving it, we present a sufficient condition for its feasibility. When the equivalent channel $\left(\mathbf{G}^{H} \mathbf{H}_{\mathrm{r}}+\mathbf{H}_{\mathrm{d}}\right)$ is full rank where $\mathbf{H}_{\mathrm{d}}=\left[\mathbf{h}_{\mathrm{d}, 1}, \cdots, \mathbf{h}_{\mathrm{d}, K}\right] \in \mathbb{C}^{N \times K}$ and $\mathbf{H}_{\mathrm{r}}=\left[\mathbf{h}_{\mathrm{r}, 1}, \cdots, \mathbf{h}_{\mathrm{r}, K}\right] \in \mathbb{C}^{M \times K}$,
%, this problem is classical feasibility characterization problem which has been extensively studied [r1], [r2]. When IRS is present, the feasibility condition is an open issue.
%In this case
problem (47) is feasible for any finite user SINR requirement $\gamma_{k}$.
The specific proof is shown in \textit{Proposition 1} in [20]. Note that this full rank
assumption is widely adopted in the literature [5], [12], [18], [20]. In fact, the channel coefficient matrix always satisfies full rank, when it independently follows some identical continuous distribution, e.g. Rayleigh
or Ricing. We also assume that the channels are full rank in our manuscript.

%\vspace{-0.6 cm}
\subsection{Beamforming Design}

When the RIS phase shifts $\boldsymbol{\Phi}_{\mathrm{r}},\boldsymbol{\Phi}_{\mathrm{t}}$, amplification gain $\mathbf{A}$  and amplitude coefficients $\bm{\varsigma}$ are fixed, the overall channel vector is determined. The sub-problem for optimizing beamforming $\mathbf{w}_{k}$ is given by
\begin{subequations}  \label{p:Minw}
\begin{align}
\min_{\mathbf{w}_{k}}~~
&{\alpha\sum_{k= 1}^{K} \left\|\mathbf{w}_{k}\right\|^{2}+(1-\alpha)\sum_{k= 1}^{K}\left\|\mathbf{A}\mathbf{G}\mathbf{w}_{k}\right\|^{2}}\\
\mathrm{s.t.}~~ &\operatorname{SINR}_{k} \geq \gamma_k, ~~k\in\mathcal{K},\label{eq:wsinr}\\
&a_m(\sum_{k=1}^{K}|\mathbf{g}_m^H\mathbf{w}_k|^2+\sigma_v^2)\leq p_{\mathrm{max},m}, ~~\forall m,
\end{align}
\end{subequations}
which is an SOCP optimization problem and can be solved by standard convex tools, e.g., the CVX solver \cite{cvx}.
%Although the constraint \eqref{eq:wsinr} is non-convex, we can perform a simple transformation on it.
%\begin{equation}
%\begin{aligned}
%\sqrt{\left(1+\frac{1}{\gamma_k}\right)}\Re\left\{\widetilde{\mathbf{h}}_{k}^{H}\mathbf{w}_{k}\right\} \geq \left\|[\widetilde{\mathbf{h}}_{k}^{H} \mathbf{W},\tilde{\sigma}_{k,\mathrm{r}}]\right\|, k \in \mathcal{K}_\mathrm{r},\\
%\sqrt{\left(1+\frac{1}{\gamma_k}\right)}\Re\left\{\widetilde{\mathbf{t}}_{k}^{H}\mathbf{w}_{k}\right\} \geq \left\|[\widetilde{\mathbf{t}}_{k}^{H} \mathbf{W},\tilde{\sigma}_{k,\mathrm{t}}]\right\|, k \in \mathcal{K}_\mathrm{t},
%\end{aligned}
%\end{equation}
%where $\tilde{\sigma}_{k,\mathrm{r}}\triangleq \sigma_{v}^{2}\|\mathbf{h}_{\mathrm{r},k}^{H}\mathbf{E}_{\mathrm{r}}\mathbf{A}\|^2+{\sigma_{k}^{2}}, k \in \mathcal{K}_\mathrm{r},$ and $\tilde{\sigma}_{k,\mathrm{t}}\triangleq \sigma_{v}^{2}\|\mathbf{h}_{\mathrm{r},k}^{H}\mathbf{E}_{\mathrm{t}}\mathbf{A}\|^2+{\sigma_{k}^{2}}, k \in \mathcal{K}_\mathrm{t},$ denote the equivalent noise received at the user. Now, the constraint \eqref{eq:wsinr} is convex and problem \eqref{p:Minw} can be solved by standard convex tools, e.g., the CVX solver \cite{cvx}.

\subsection{Amplification Gain Design}

When other variables are fixed, the optimization problem of amplification matrix $\mathbf{A}$ can be expressed by
\begin{subequations}  \label{p:Minp}
\begin{align}
\min_{\mathbf{A}}~~
&{(1-\alpha)\sum_{k= 1}^{K}\left\|\mathbf{A}\mathbf{G}\mathbf{w}_{k}\right\|^{2}}\label{eq:Minpowerp}\\
\mathrm{s.t.}~~ &\operatorname{SINR}_{k} \geq \gamma_k, ~~k\in\mathcal{K},\\
&a_m(\sum_{k=1}^{K}|\mathbf{g}_m^H\mathbf{w}_k|^2+\sigma_v^2)\leq p_{\mathrm{max},m}, ~~\forall m,
\end{align}
\end{subequations}
which is again an SOCP problem and can be efficiently solved.

\subsection{Phase Shifts Design}
After obtaining the BS beamforming $\mathbf{w}_{k}, \forall k,$ and RIS amplification gain $\mathbf{A}$, the objective of the original optimization problem \eqref{p:Minpower} has been determined. This means that, with
given $\mathbf{w}_{k}$ and $\mathbf{A}$, the design of phase-shift matrices $\boldsymbol{\Phi}_{\mathrm{r}},\boldsymbol{\Phi}_{\mathrm{t}}$ and amplitude coefficients $\bm{\varsigma}$ becomes a feasibility-check problem and could not directly affect the power minimization objective of \eqref{p:Minpower}. Therefore, we attempt to formulate another proper objective function that can facilitate the reduction of the total power consumption for next iteration and guarantee its feasibility.

We note that the conditionally optimal $\mathbf{w}_{k}$ and $\mathbf{A}$ of the power minimization problem \eqref{p:Minpower} usually make constraint \eqref{eq:sinr} almost equal, i.e., the QoS requirement is satisfied almost with equality. In order to further reduce the total power consumption in the next iteration, we propose to use the QoS balancing as the objective function, which can introduce an improved QoS and provide more freedom for power minimization in the next iteration.
To this end, when optimizing phase-shift vector $\boldsymbol{\phi}_{\mathrm{r}}$, the QoS balancing problem can be formulated as
\begin{subequations}  \label{p:sinr_pr}
\begin{align}
\max _{\bm{\phi}_{\mathrm{r}}}&\min_{k\in\mathcal{K}_{\mathrm{r}}} ~\frac{\operatorname{SINR}_{k}}{\gamma_k} \label{eq:sinrfpr}\\
\mathrm{s.t.}& ~~
|\bm{\phi}_{\mathrm{r},m}|=1,~~\forall m. \label{eq:sinrprunit}
\end{align}
\end{subequations}
Unfortunately, problem \eqref{p:sinr_pr} is very challenging to handle due to the following reasons: \textit{i}) the objective function \eqref{eq:sinrfpr} is max-min function and non-differentiable which hamper the algorithm development, \textit{ii}) the fractional terms in objective function \eqref{eq:sinrfpr} are non-convex, \textit{iii}) the unit modulus constraint for the RIS phase-shifts in \eqref{eq:sinrprunit} is also non-convex. In order to tackle these difficulties, the max-min function is equivalently transformed into min-max form to apply the log-sum-exp approach and Riemannian-manifold algorithm. Then, we reformulate the fractional terms in \eqref{eq:sinrfpr} into a form with decoupled numerator and denominator by applying Dinkelbach's transform \cite{dk}. Therefore, by introducing a new auxiliary variable $\varpi_{\mathrm{r}}$, the transformed problem can be expressed as
\begin{subequations}  \label{p:resinr}
\begin{align}
\min_{\bm{\phi}_{\mathrm{r}},\varpi_{\mathrm{r}}}&
\max_{k\in\mathcal{K}_{\mathrm{r}}}~\left\{f_{k}-\varpi_{\mathrm{r}} g_{k}\right\} \\
\mathrm{s.t.}
& ~~|\bm{\phi}_{\mathrm{r},m}|=1, ~~\forall m,
\end{align}
\end{subequations}
where $f_{k}$ and $g_{k}$ are auxiliary functions which are presented as
\begin{subequations} \label{eq:gk}
\begin{align}
f_k &\triangleq \gamma_k(\sum_{j \neq k}^{K}\left|\left(\mathbf{h}_{\mathrm{d},k}^{H}+\mathbf{h}_{\mathrm{r},k}^{H} \bm{\Phi}_{\mathrm{r}}\mathbf{E}_{\mathrm{r}}\mathbf{A} \mathbf{G}\right)\mathbf{w}_{j} \right|^{2}\nonumber \\
&~~+
\sigma_{v}^{2}\|\mathbf{h}_{\mathrm{r},k}^{H}\mathbf{E}_{\mathrm{r}}\mathbf{A}\|^2+{\sigma_{k}^{2}}),~~k\in \mathcal{K}_\mathrm{r},
\\g_k &\triangleq\left|\left(\mathbf{h}_{\mathrm{d},k}^{H}+\mathbf{h}_{\mathrm{r},k}^{H}\bm{\Phi}_{\mathrm{r}} \mathbf{E}_{\mathrm{r}}\mathbf{A} \mathbf{G}\right)\mathbf{w}_{k}\right|^2,~~k\in \mathcal{K}_\mathrm{r}.
\end{align}
\end{subequations}
The optimal $\varpi_{\mathrm{r}}$ can be expressed in a closed-form as \cite{dk}
\begin{equation}
\begin{aligned}
\varpi_{\mathrm{r}}^\star = \max\left\{\frac{f_{k}}{g_{k}}\right\},~~\forall k\in \mathcal{K}_\mathrm{r}.
\end{aligned}
\end{equation}
However, problem \eqref{p:resinr} is still non-differentiable. Then we smooth the max function by the well-known log-sum-exp approximation and obtain
\begin{subequations}  \label{p:lastsinr}
\begin{align}
\min_{\bm{\phi}_{\mathrm{r}},\varpi_{\mathrm{r}}}~
& \varepsilon \log \sum_{k\in \mathcal{K}_\mathrm{r}} \exp \left\{\frac{f_{k}-\varpi_{\mathrm{r}} g_{k}}{\varepsilon}\right\} \\
\mathrm{s.t.} ~~& |\bm{\phi}_{\mathrm{r},m}|=1, ~~\forall m,\label{eq:lastsinrunit}
\end{align}
\end{subequations}
where $\varepsilon$ is a relatively small positive number to maintain the approximation. While the objective of \eqref{p:lastsinr} is smooth and differentiable, the non-convex unit modulus constraint \eqref{eq:lastsinrunit} still makes the problem difficult to solve. Therefore, we use Riemannian-manifold algorithm presented in Section \ref{s:sumrate} to solve \eqref{p:lastsinr}. Its Euclidean gradient can be derived by
\begin{equation}
\begin{aligned}
\mathbf{g}_1 \triangleq \frac{\sum_{k\in \mathcal{K}_\mathrm{r}}\exp \left\{\frac{f_{k}-\varpi_{\mathrm{r}} g_{k}}{\varepsilon}\right\}\left(\mathbf{a}_{k}-\varpi_{\mathrm{r}}\mathbf{b}_{k}\right)}{\sum_{k\in \mathcal{K}_\mathrm{r}} \exp \left\{\frac{f_{k}-\varpi_\mathrm{r} g_{k}}{\varepsilon}\right\}},~~k\in \mathcal{K}_\mathrm{r},
\end{aligned}
\end{equation}
where we define
\begin{subequations}
\begin{align}
\mathbf{a}_{k}&\triangleq 2\gamma_k\sum_{j \neq k}^{K}(\mathbf{r}_{k,j}\mathbf{r}_{k,j}^H\bm{\phi}_{\mathrm{r}}+\mathbf{r}_{k,k}\mathbf{h}_{\mathrm{d},k}^H\mathbf{w}_{j}),~~k\in \mathcal{K}_\mathrm{r},\\
\mathbf{b}_{k}&\triangleq 2(\mathbf{r}_{k,k}\mathbf{r}_{k,k}^H\bm{\phi}_{\mathrm{r}}
+\mathbf{r}_{k,k}\mathbf{h}_{\mathrm{d},k}^H\mathbf{w}_{k}),~~k\in \mathcal{K}_\mathrm{r}.
\end{align}
\end{subequations}
Then, the Riemannian-manifold-based phase-shift design presented in Algorithm \ref{alg:Riemannian} can be applied.

Similarly, as we discussed above, the optimization problem of $\boldsymbol{\phi}_{\mathrm{t}}$ can be expressed by
\begin{subequations}  \label{p:lastsinrt}
\begin{align}
\min_{\bm{\phi}_{\mathrm{t}},\varpi_{\mathrm{t}}}~
& \varepsilon \log \sum_{k\in \mathcal{K}_\mathrm{t}} \exp \left\{\frac{f_{k}-\varpi_{\mathrm{t}} g_{k}}{\varepsilon}\right\} \\
\mathrm{s.t.} ~~& |\bm{\phi}_{\mathrm{t},m}|=1, ~~\forall m,\label{eq:lastsinrunitt}
\end{align}
\end{subequations}
where $f_{k}$ and $g_{k}$ are presented as
\begin{subequations} \label{eq:fk}
\begin{align}
f_k &\triangleq  \gamma_k(\sum_{j \neq k}^{K}\left|\left(\mathbf{h}_{\mathrm{d},k}^{H}+\mathbf{h}_{\mathrm{r},k}^{H} \bm{\Phi}_{\mathrm{t}}\mathbf{E}_{\mathrm{t}}\mathbf{A} \mathbf{G}\right)\mathbf{w}_{j} \right|^{2}\nonumber \\
&~~+
\sigma_{v}^{2}\|\mathbf{h}_{\mathrm{r},k}^{H}\mathbf{E}_{\mathrm{t}}\mathbf{A}\|^2+{\sigma_{k}^{2}}),~~k\in \mathcal{K}_\mathrm{t},\\
g_k &\triangleq\left|\left(\mathbf{h}_{\mathrm{d},k}^{H}+\mathbf{h}_{\mathrm{r},k}^{H}\bm{\Phi}_{\mathrm{t}} \mathbf{E}_{\mathrm{t}}\mathbf{A} \mathbf{G}\right)\mathbf{w}_{k}\right|^2,~~k\in \mathcal{K}_\mathrm{t}.
\end{align}
\end{subequations}
The optimal $\varpi_{\mathrm{t}}$ can be expressed as \cite{dk}
\begin{equation}
\begin{aligned}
\varpi_{\mathrm{t}}^\star = \max\left\{\frac{f_{k}}{g_{k}}\right\},~~\forall k\in \mathcal{K}_\mathrm{t}.
\end{aligned}
\end{equation}
Likewise, the Euclidean gradient of problem \eqref{p:lastsinrt} can be derived by
\begin{equation}
\begin{aligned}
\mathbf{g}_2 \triangleq \frac{\sum_{k\in \mathcal{K}_\mathrm{t}}\exp \left\{\frac{f_{k}-\varpi_{\mathrm{t}} g_{k}}{\varepsilon}\right\}\left(\mathbf{c}_{k}-\varpi_{\mathrm{t}}\mathbf{d}_{k}\right)}{\sum_{k\in \mathcal{K}_\mathrm{t}} \exp \left\{\frac{f_{k}-\varpi_\mathrm{t} g_{k}}{\varepsilon}\right\}},~~k\in \mathcal{K}_\mathrm{t},
\end{aligned}
\end{equation}
where
\begin{subequations}
\begin{align}
\mathbf{c}_{k}\triangleq ~&2\gamma_k\sum_{j \neq k}^{K}(\mathbf{r}_{k,j}\mathbf{r}_{k,j}^H\bm{\phi}_{\mathrm{t}}+\mathbf{r}_{k,k}\mathbf{h}_{\mathrm{d},k}^H\mathbf{w}_{j}),~~k\in \mathcal{K}_\mathrm{t},\\
\mathbf{d}_{k}\triangleq ~&2(\mathbf{r}_{k,k}\mathbf{r}_{k,k}^H\bm{\phi}_{\mathrm{t}}
+\mathbf{r}_{k,k}\mathbf{h}_{\mathrm{d},k}^H\mathbf{w}_{k}),~~k\in \mathcal{K}_\mathrm{t}.
\end{align}
\end{subequations}
Therefore, the phase-shift $\bm{\phi}_{\mathrm{t}}$ can be obtained by the Riemannian-manifold-based algorithm in Algorithm \ref{alg:Riemannian}.
Moreover, since the RIS phase-shift $\boldsymbol{\phi}_{\mathrm{r}}$ and $\boldsymbol{\phi}_{\mathrm{t}}$ are separate, they can be optimized in parallel to reduce computational time.

{
\subsection{Amplitude Coefficients Design}}

Given other variables, the optimization problem of amplitude coefficients $\bm{\varsigma}$ can be reformulated as follows
\begin{subequations}  \label{p:minvar}
\begin{align}
\min_{\varsigma_m}~~
& \varepsilon \log \sum_{k\in \mathcal{K}} \exp \left\{\frac{f_{k}-\varpi g_{k}}{\varepsilon}\right\}\\
\mathrm{s.t.} ~~& \varsigma_m\in[0,1], ~~\forall m.
\end{align}
\end{subequations}
where $\varpi^\star= \frac{f_{k}}{g_{k}}, \forall k \in \mathcal{K}$.
As we discussed above, this problem is a constrained optimization problem with one variable ${\varsigma}_m$ and we iteratively design each element of the vector $\bm{\varsigma}$, e.g., ${\varsigma}_m$ until convergence.

\label{alg:minpower}
\begin{algorithm}[!t]\begin{small}
\caption{Joint BS Beamforming and RIS Design for the Power Minimization Problem}
\label{alg:PM}
\begin{algorithmic}[1]
\REQUIRE $\mathbf{h}_{\mathrm{d},k}^H$, $\mathbf{h}_{\mathrm{r},k}^H$, $\mathbf{G}$, $\sigma^2$, $\sigma_v^2$, $\operatorname{SINR}_{k}, \forall k$, $T_\text{max}$, $\delta_\text{th}$.
\ENSURE $\mathbf{w}_k^\star$, $\bm{\Phi}_{\mathrm{r}}^\star, \bm{\Phi}_{\mathrm{t}}^\star$, $\mathbf{A}^\star$ and $\bm{\varsigma}^\star$.
    \STATE {Initialize $\mathbf{w}_k$, $\bm{\Phi}_{\mathrm{r}}, \bm{\Phi}_{\mathrm{t}}$, $\mathbf{A}$, $\bm{\varsigma}$, $t = 1$, $\delta = \infty$, $p_{\text{temp}}=0$.}
    \STATE{\textbf{while} $t\leq T_\text{max}$ and $\delta \geq \delta_\text{th}$ \textbf{do}}
        \STATE{\hspace{0.4 cm}$p_{\text{pre}} = p_{\text{temp}}$.}
        \STATE{\hspace{0.4 cm}Update BS beamforming $\mathbf{w}_k^\star, \forall k$ by (\ref{p:Minw}).}
        \STATE{\hspace{0.4 cm}Update RIS amplification gain $\mathbf{A}^\star$ by (\ref{p:Minp}).}
        \STATE{\hspace{0.4 cm}Update reflection matrices $\bm{\Phi}_{\mathrm{r}}^\star, \bm{\Phi}_{\mathrm{t}}^\star$ using Algorithm \ref{alg:Riemannian}.}
        \STATE{\hspace{0.4 cm}{Update amplitude coefficients $\bm{\varsigma}^\star$ using Algorithm \ref{alg:ratio}.}}
        \STATE{\hspace{0.4 cm}$p_{\text{temp}} = \sum_{k= 1}^{K} \left\|\mathbf{w}_{k}\right\|^{2}+\sum_{k= 1}^{K}\left\|\mathbf{A}\mathbf{G}\mathbf{w}_{k}\right\|^{2}$.}
        \STATE{\hspace{0.4 cm}$\delta = \frac{\left|p_{\text{temp}} - p_{\text{pre}}\right|}{p_{\text{temp}}}$.}
        \STATE{\hspace{0.4 cm}$t=t+1$.}
    \STATE{\textbf{end while}}
    \end{algorithmic}\end{small}
\end{algorithm}

%\subsection{Power Ratio Design}

%The optimization problem of power ratio $\bm{\varsigma}$ can be reformulated as follows
%\begin{subequations}  \label{p:minvar}
%\begin{align}
%\min_{\varsigma_m}~~
%& \varepsilon \log \sum_{k=1}^{K} \exp \left\{\frac{f_{k}-\varpi g_{k}}{\varepsilon}\right\}\\
%\mathrm{s.t.} ~~& \varsigma_m\in[0,1], ~~~~\forall m.
%\end{align}
%\end{subequations}
%As mentioned above, this problem is a constrained optimization problem with one variable $\bm{\varsigma}$. We iteratively design each element of the vector $\bm{\varsigma}$ until convergence.

Finally, the joint BS beamforming and dual-functional RIS design for the total power minimization problem is straightforward. Similarly, with a proper initialization, the BS transmit beamforming $\mathbf{w}_{k}$, amplification gain matrix $\mathbf{A}$, phase-shift matrices $\bm{\Phi}_{\mathrm{r}},\bm{\Phi}_{\mathrm{t}}$ and amplitude coefficients $\bm{\varsigma}$ are iteratively optimized  until convergence, which is summarized in Algorithm \ref{alg:PM}.

%\vspace{-0.3 cm}

\subsection{Complexity Analysis}
We also give a brief complexity analysis of the proposed joint BS beamforming and RIS design for the power minimization problem. The optimizations of the BS transmit beamforming $\mathbf{w}_k$ and RIS amplification gain matrix $\mathbf{A}$ are SOCP problems and have the complexity of approximately $\mathcal{O}((KN)^{3}(M+K)^{1.5})$ and $\mathcal{O}(M^{3}(M+K)^{1.5})$.
{Then the computational complexity of phase-shift design based on Riemannian-manifold and Dinkelbach's transform is approximately at most $\mathcal{O}(I_{d1}M^{2})$ and $\mathcal{O}(I_{d2}M^{2})$,} wherein the parameters $I_{d1}$ and $I_{d2}$ are the number of iterations of updating phase shift $\bm{\phi}_\mathrm{r}$ and $\bm{\phi}_\mathrm{t}$. Hence, the total computational complexity of Algorithm 4 is approximated by $\mathcal{O}(I_P((KN)^{3}(M+K)^{1.5}+M^{3}(M+K)^{1.5}+I_{d1}M^{2}+I_{d2}M^{2}+I_\varsigma M))$, where $I_P$ denotes the required iteration number for convergence.

\vspace{-0.0 cm}
\section{Simulation Results}
\label{s:simulation}
In this section, we provide extensive simulation results to demonstrate the advantages of the proposed dual-functional active RIS architecture and illustrate the effectiveness of our proposed algorithms. For simplicity, we assume the QoS requirement and the noise power at all receivers are the same,
i.e., $\gamma_k=\gamma, \sigma_k^2 = -80$ dBm, $\forall k$.
The factor $\varepsilon$ to maintain the approximation is set as $10^{-3}$.
%The maximum transmit power at BS and the maximum amplifing power at active RIS are set as $P_\mathrm{T} = $Pmax BS = PmaxA = 10 mW
The transmit antenna array at the BS is assumed to be a uniform linear array with antenna spacing given by $\lambda/2$ where $\lambda$ denotes the wavelength. The thermal noise power introduced by active RIS  is $\sigma_v^2 = -80$dBm.
The distance-dependent path loss is modeled as $\text{PL}(d) = C_0\left(d_0/d\right)^\kappa$, where $C_0 = -30$dB is the path loss for the reference distance $d_0 = 1$m, $d$ is the link distance, and $\kappa$ is the path-loss exponent.

\begin{figure}[t]
\centering
\includegraphics[height=2.02 in]{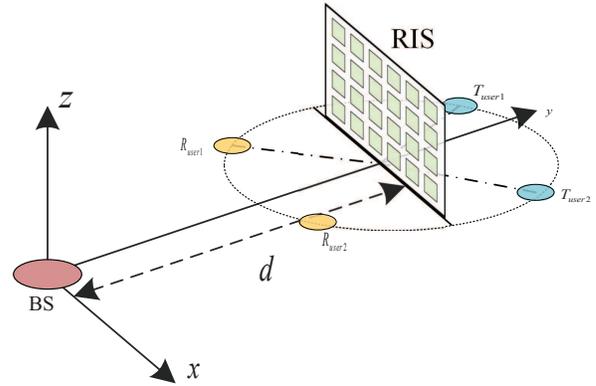}
\caption{Simulation setup for MU-MISO system.}\label{fig:setting}
\vspace{0.0 cm}
\end{figure}

In addition, the BS-RIS channel is assumed to follow the small-scale Rician fading channel model, which consists of line-of-sight (LoS) and non-LoS (NLoS) components. The channels from the BS to the RIS can be expressed as
\begin{equation}
\mathbf{G}=\sqrt{\frac{\alpha_g}{\alpha_g+1}} \mathbf{G}^{\operatorname{LoS}}+\sqrt{\frac{1}{\alpha_g+1}} \mathbf{G}^{\mathrm{NLoS}},
\end{equation}
where $\alpha_g$ is the Rician factor set as 3dB, $\mathbf{G}^{\operatorname{LoS}}$ is the LoS component which depends on the geometric settings, and $\mathbf{G}^{\mathrm{NLoS}}$ is the NLoS Rayleigh fading component with a path-loss exponent of 2.5.
The channels from the RIS to users only have NLoS components and the path-loss exponent is 2.0, while the path-loss exponent of channels from the BS to users in $\mathcal{K}_\mathrm{t}$ is 3.6. Since the channels from BS to users in $\mathcal{K}_\mathrm{t}$ that are located at the back of RIS are weak, due to the blockage and severe path loss, the path-loss exponent is assumed to be 4.2 usually.
The BS is equipped with $N = 16$ antennas and serves $K = 4$ users with two users in $\mathcal{K}_\mathrm{r}$ and two users in $\mathcal{K}_\mathrm{t}$.
We assume the active RIS equipped with $M = 128$ elements is 80m away from the BS ($d= 80$m), while the users are randomly deployed on circle centered at the active RIS with the radius of 10m ($r_\mathrm{u}$ = 10m), as illustrated in Fig. \ref{fig:setting}.
The simulation results were obtained by averaging over $10^{4}$ random channel realizations.
\begin{figure}[t]
\centering
\includegraphics[height=2.74 in]{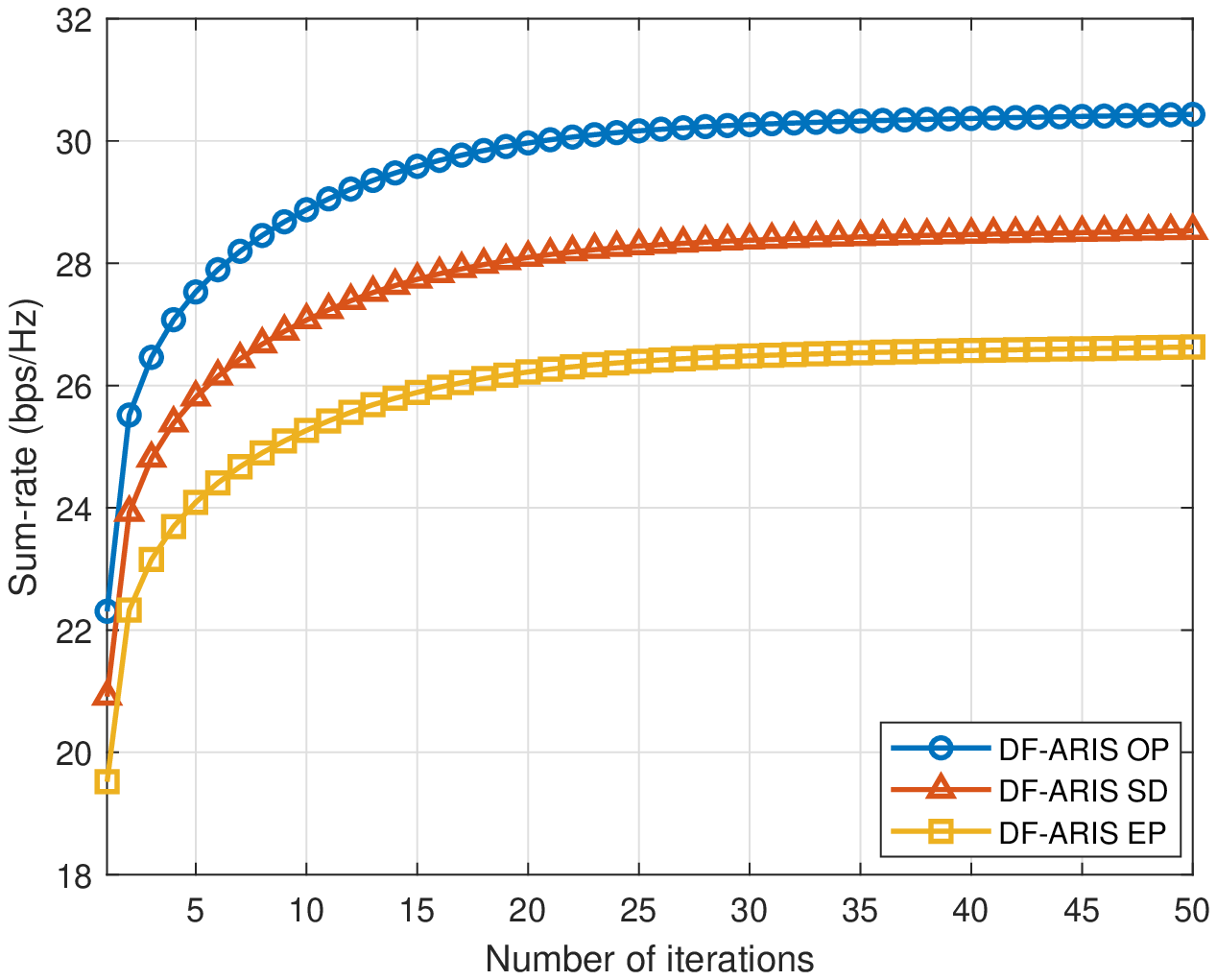}
\caption{Sum-rate versus the number of iterations ($P_\mathrm{T}$ = 16dBm, $P_\mathrm{R}$ = 10dBm, $M$ = 128, $K$ = 4).}\label{fig:convergence}
%\vspace{-0.2 cm}
\end{figure}

\begin{figure}[t]
\centering
\includegraphics[height=2.74 in]{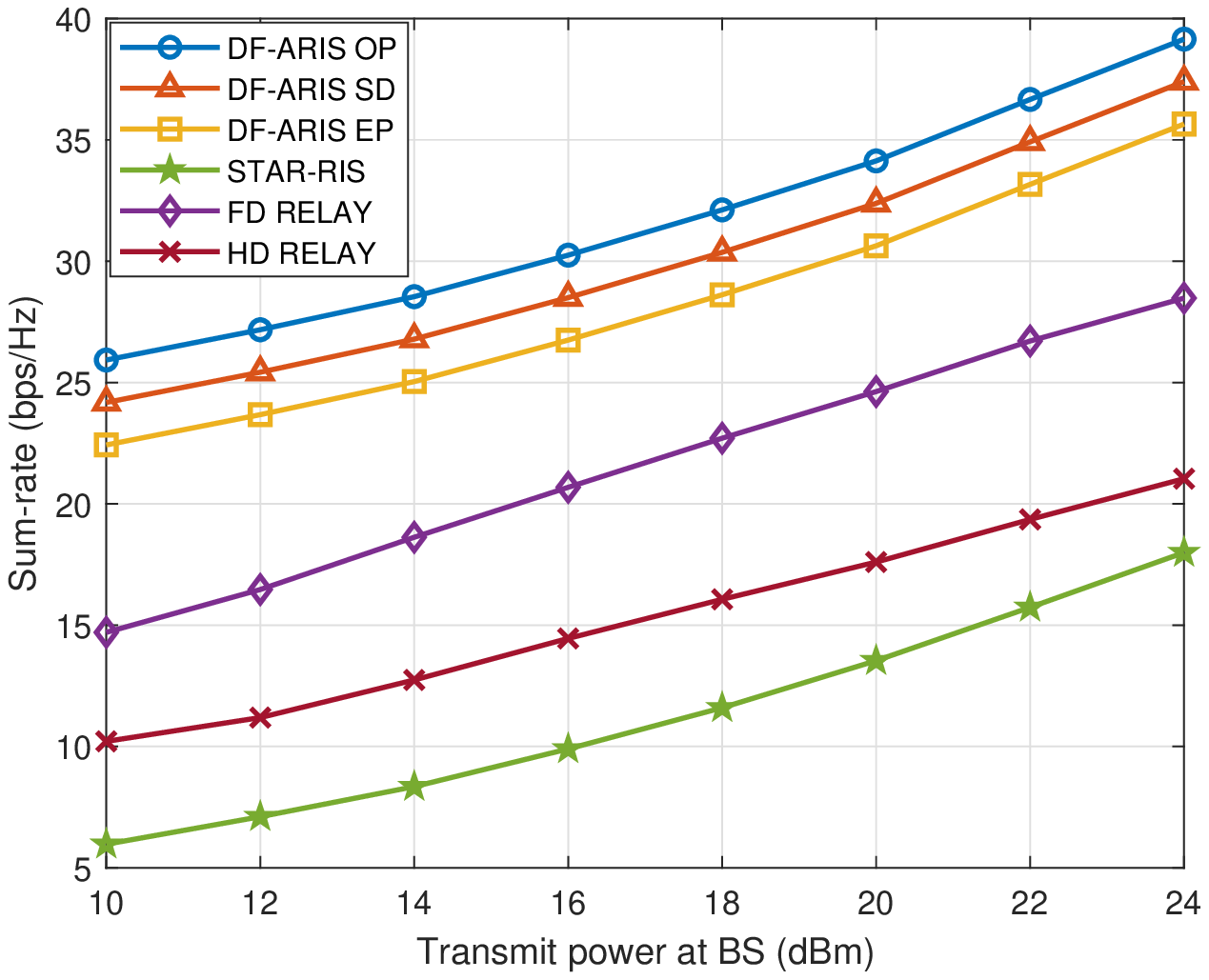}
\caption{Sum-rate versus the transmit power at BS $P_\mathrm{T}$ ($P_\mathrm{R}$ = 10dBm, $M$ = 128, $K$ = 4).}\label{fig:ratevsp}
\vspace{0.1 cm}
\end{figure}
\begin{figure}[!t]
\centering
\includegraphics[height=2.74 in]{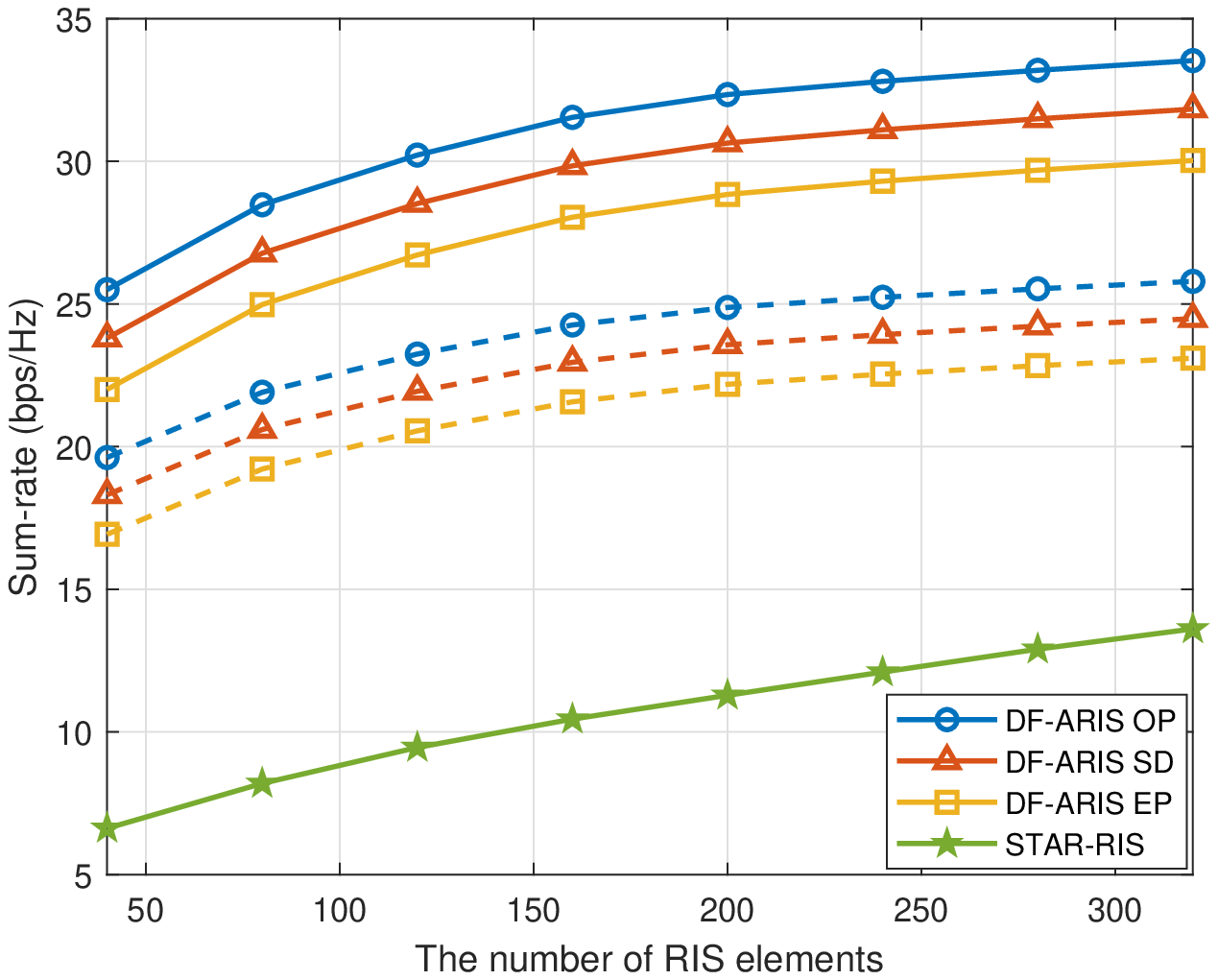}
\caption{Sum-rate versus the number of RIS elements $M$ ($P_\mathrm{T}$ = 16dBm, $K$ = 4. The solid lines indicate $P_\mathrm{R}$ = 10dBm and the dashed lines indicate $P_\mathrm{R}$ = 0dBm).}\label{fig:ratevsm}
\vspace{-0.3 cm}
\end{figure}

We also include two other operation modes for the active RIS: \textit{i}) Space Division (SD) mode, i.e., elements are divided into two groups, one group operates for the reflection functionality while the other group operates for the transmission functionality. In this mode, {the amplitude coefficients $\varsigma_m$ is fixed as 1 or 0 for these two groups, respectively.} \textit{ii}) Equal Power (EP) mode, i.e., the powers of reflected signal and transmitted signal are fixed at the same value as $\varsigma_m = {1}/{\sqrt{2}}$. Algorithms for optimising other variables also apply to these modes.

\subsection{Sum-Rate Maximization}

In this subsection, we present simulation results for the sum-rate maximization problem.
Fig. \ref{fig:convergence} shows the convergence of our proposed algorithm for the cases that the dual-functional active RIS (DF-ARIS) has different operation modes: \textit{i}) Optimal Power (OP) mode, i.e, the amplitude coefficients $\varsigma_m$ can be continuously selected within the range $[0,1],$ \textit{ii}) Space Division (SD) mode, \textit{iii}) Equal Power (EP) mode.
It can be observed that our proposed algorithm converges within 40 iterations.

In Fig. \ref{fig:ratevsp}, we show the sum-rate versus the transmit power $P_\mathrm{T}$. For comparison purpose, we also include:
%\textit{i}) The active RIS with Space Division (SD) mode, i.e., elements are divided into two groups, one group operates for the reflection functionality while the other group operates for the transmission functionality. In this SD mode, the power ratio $\varsigma_m$ is fixed as 1 or 0 for these two groups, respectively. \textit{ii}) The active RIS with Equal Power (EP) mode, i.e., the powers of reflected signal and transmitted signal are fixed at the same value as $\varsigma_m = {1}/{\sqrt{2}}$. The transmit beamforming and RIS phase-shift designs for SD mode and EP mode also utilize the proposed algorithm framework with corresponding fixed $\varsigma_m$. Moreover, in order to demonstrate the effectiveness of our proposed dual-functional active RIS architecture, we also consider
\textit{i}) traditional full-duplex relay (FD RELAY) with 8 antennas \cite{comprelay}; \textit{ii}) half-duplex relay (HD RELAY) with 8 antennas \cite{comprelay}, which has the same power budget as the active RIS and FD Relay; \textit{iii}) passive STAR-RIS scheme \cite{STAR1}, in which the power consumption at the BS is the same with the total power consumption of active RIS-aided system for fair comparison.
For fair comparison, the traditional passive RIS is excluded in the following simulations since it cannot effectively serve the users at the back side of it. Instead, the passive STAR-RIS can be considered as a more advanced version of the traditional RIS for providing full coverage. The sum-rate of all schemes increases with larger transmit power. Moreover, the sum-rate achieved by our proposed dual-functional active RIS architecture is dramatically greater than the competitor in \cite{STAR1}, which is a passive device without signal amplification. In addition, our proposed dual-functional active RIS also outperforms the  traditional full-duplex and half-duplex relay owing to the substantial beamforming gain provided by the RIS. These results support the benefit of deploying active RIS. Since the degrees of freedom of SD and EP modes are limited, the sum-rates achieved by these two schemes are lower than optimal power mode, which demonstrates the effectiveness of power splitting.

\begin{figure}[t]
\centering
\includegraphics[height=2.74 in]{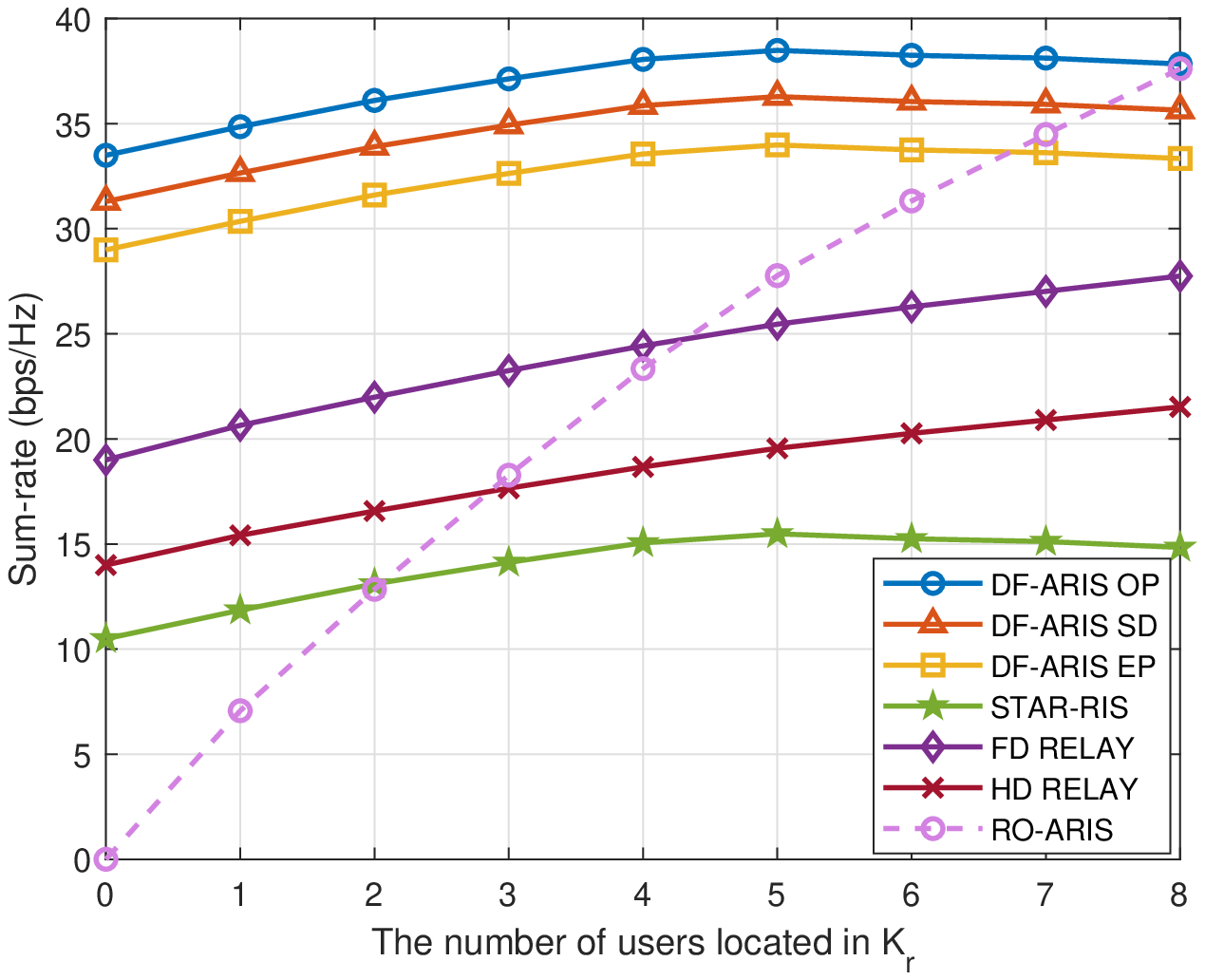}
\caption{Sum-rate versus the number of users in $\mathcal{K}_\mathrm{r}$ ($P_\mathrm{T}$ = 16dBm, $P_\mathrm{R}$ = 10dBm, $M$ = 128, $K$ = 8).}\label{fig:rate_user}
\vspace{0.0 cm}
\end{figure}

%\begin{figure}[t]
%\centering
%\includegraphics[height=2.7 in]{ratevsratio.eps}
%\caption{Sum-rate versus power ratio of the reflected and transmitted signals ($P_\mathrm{T}$ = 16dBm, $P_\mathrm{R}$ = 10dBm, $M$ = 256).}\label{fig:ratevsratio}
%\vspace{0.0 cm}
%\end{figure}
%
%In Fig. \ref{fig:ratevsratio}, we analyze the relation between the sum-rate and the power ratio of the
%reflected and transmitted signals $\bm{\varsigma}$ and we can note that the sum-rate of all users is maximized for $\varsigma_m = 0.5$. Moreover, we can see that the sum-rate of the users in $\mathcal{K}_\mathrm{t}$ is zero when $\varsigma_m = 1$, which is because it actually performs as a reflective RIS in this case. However, when the dual-functional RIS acts as a full transmissive RIS, e.g., $\varsigma_m = 0$, the sum-rate of the users in $\mathcal{K}_\mathrm{r}$ is not zero due to the direct links from BS to them.

Next, we show the achievable sum-rate versus the number of RIS elements $M$ in Fig. \ref{fig:ratevsm}.
Since the larger RIS can provide larger beamforming gain, we observe that the sum-rate increases for all schemes with increasing $M$. Moreover, our proposed scheme always achieves significantly better sum-rate performance with different RIS sizes.

\begin{figure}[t]
\centering
\includegraphics[height=2.74 in]{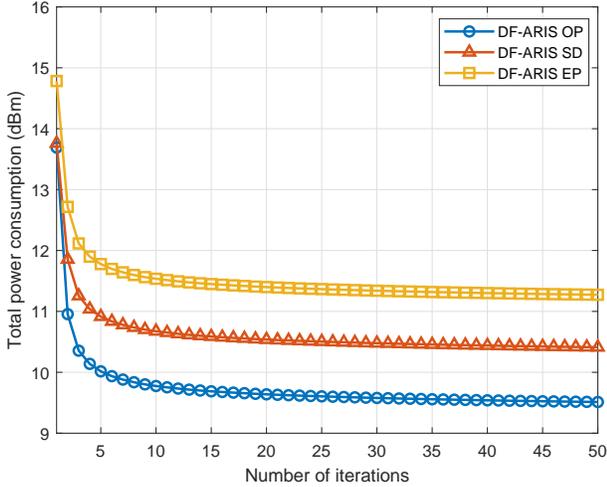}
\caption{Total power consumption versus the number of iterations ($\gamma$ = 12dB, $M$ = 128, $K$ = 4).}\label{fig:p_convergence}
\vspace{-0.2 cm}
\end{figure}
\begin{figure}[t]
\centering
\includegraphics[height=2.74 in]{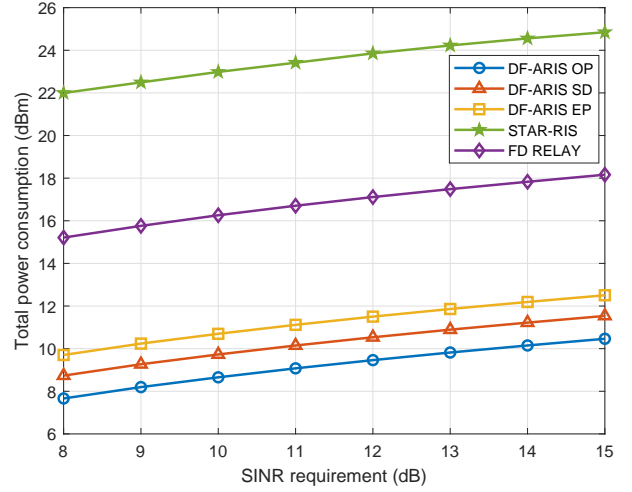}
\caption{Total power consumption versus SINR requirement ($M$ = 128, $K$ = 4).}\label{fig:PvsSINR}
\vspace{-0.1 cm}
\end{figure}
\begin{figure}[t]
\centering
\includegraphics[height=2.74 in]{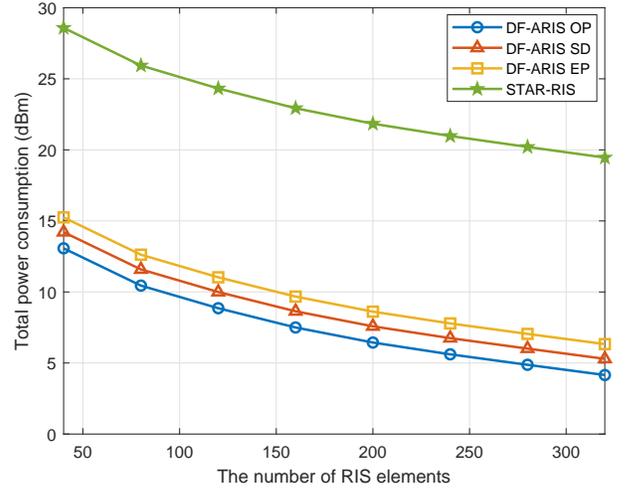}
\caption{Total power consumption versus the number of RIS elements ($\gamma$ = 12dB, $K$ = 4).}\label{fig:pvsm}
%\vspace{-0.35 cm}
\end{figure}
\begin{figure}[t]
\centering
\includegraphics[height=2.74 in]{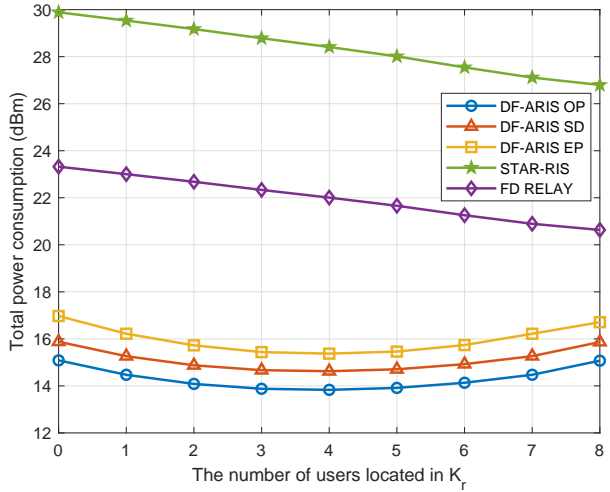}
\caption{Total power consumption versus the number of users in $\mathcal{K}_\mathrm{r}$ ($\gamma$ = 12dB, $M$ = 128, $K$ = 8).}\label{fig:power_user}
\vspace{0.0 cm}
\end{figure}

In order to demonstrate the impact of distribution of users with respect to the location of the dual-functional
active RIS, in Fig. \ref{fig:rate_user} we compare the sum-rate versus the number of users in $\mathcal{K}_\mathrm{r}$, in which
the users are randomly located on circle centered at the active RIS. We can see from Fig. \ref{fig:rate_user} that the sum-rate is maximized when users are near equally distributed on the two sides of the RIS. For
comparison, we also investigate the performance of current reflective-only active RIS (RO-ARIS) proposed in \cite{Active}, \cite{liang}. It is noted that, in RO-ARIS assisted communication system,
the sum-rate increases linearly with the number of users in $\mathcal{K}_\mathrm{r}$, since it can only assist the
communication of the users within the same half-space as BS. However, the proposed dual-functional RIS can always provide satisfactory sum-rate performance for different user distributions.

%In order to demonstrate the impact of RIS location, in Fig. \ref{fig:ratevsd}, we compare the sum-rate versus the location of RIS, in which the sum of the distances between BS and RIS, RIS and users is fixed as 220m. The distance between BS and RIS is varying from 60m to 180m.
%%In general, the active RIS has a better performance than STAR.
%%It is also observed that when the RIS has high power, it is better to deploy it closer to the users, i.e., the maximum sum-rate achieved when $P_\mathrm{R}$ = 10dBm is at 180m while the maximum sum-rate achieved when $P_\mathrm{R}$ = 0dBm is at 140m.
%We can see from Fig. \ref{fig:ratevsd} that the STAR-RIS has the worst performance when the it is located close to the middle point between the BS and users (i.e., d = 120m), where the reflection channel become the weakest due to the double-fading effect. However, the active RIS can effectively overcome the double-fading attenuation and always provide satisfactory sum-rate performance for different RIS locations.
%%The sum-rate increases when the active RIS comes closer to the users, where the active RIS can provide more amplification gain to compensate for the attenuation caused by the double-fading path loss.

\subsection{Power Minimization}
In this subsection, we illustrate the simulation results for the power minimization problem.
{The power consumption weighted factor $\alpha$ is set as 0.5.}
The convergence performance shown in Fig. \ref{fig:p_convergence} is similar to that observed for the sum-rate maximization problem in Fig. \ref{fig:convergence}. It can be seen from Fig. \ref{fig:p_convergence} that all schemes converge around 40 iterations.
%We start with presenting the convergence of the proposed joint BS beamforming and RIS reflection and transmission design by plotting the total power consumption versus the number of iterations in Fig. \ref{fig:conv}. Simulation results illustrate that the proposed algorithm can converge within 15 iterations. For the case of employing low-resolution ($b=1,2,3$ bit) phase shifters, the proposed algorithm will converge faster within 10 iterations.

In Fig. \ref{fig:PvsSINR}, we show the total power consumption versus the SINR requirement of users.
It can be observed that our proposed scheme requires much less total power consumption than the traditional full-duplex relay and STAR-RIS, which validates the advantage of energy efficiency of the proposed dual-functional active RIS scheme. More interestingly, the passive STAR-RIS requires dramatically larger power consumption at the BS to overcome the double fading.

Next, we present the total power consumption versus the number of RIS elements $M$ in Fig. \ref{fig:pvsm}. We observe that as the number of RIS elements increases, the total power consumption is greatly reduced. A similar conclusion can be drawn from Fig. \ref{fig:pvsm} that the proposed active RIS architecture can always achieve
better performance.

Finally, in Fig. \ref{fig:power_user} we show the total power consumption versus the number of users in $\mathcal{K}_\mathrm{r}$. The proposed dual-functional active RIS always has a lower power consumption performance than traditional FD relay and STAR-RIS. The minimum total power consumption is achieved when users are equally located at the two sides of the RIS, i.e. there are 4 users located in $\mathcal{K}_\mathrm{r}$. When all users are mostly located at the same side of the RIS, the beamforming gain provided by large-scale phase shifters on the other side is limited, which leads to an increase in power consumption.

%the power of signals re-emitted by both sides of the surface cannot be optimized for all users concurrently, which leads to an increase in power consumption.

%Finally, in Fig. \ref{fig:Pvsd} we show the total power consumption versus the location of RIS. The proposed dual-functional active RIS always has a lower power consumption performance than traditional HD relay and STAR-RIS. The better performance is achieved when the RIS moves closer to users. When the active RIS is at the middle between the BS and users, more power is required to overcome the double-fading path loss.

\section{Conclusion}
\label{s:conclusion}

In this article, we investigated a novel reflection and transmission dual-functional active RIS architecture, which can simultaneously realize reflection and transmission functionalities with signal amplification to enhance the QoS of all users and extend the coverage. In particular, we considered the joint BS beamforming and active RIS designs for MU-MISO systems. Efficient iterative algorithms were proposed to solve the sum-rate maximization and power minimization problems, respectively.
The simulation results illustrated that our proposed dual-functional RIS architecture exhibited remarkably better performance in terms of sum-rate enhancement and power-savings, which confirmed that the employment of dual-functional active RIS in wireless communication systems has promising prospects performing as a multi-antenna and low-power relay.

\end{document}